%% file: main.tex
\documentclass[trackchanges,twocolumn,times]{aastex7}

\usepackage{newtxtext,newtxmath}
\usepackage[T1]{fontenc}
\usepackage{graphicx}	
\usepackage{amsmath}	
\usepackage{float}
\usepackage{bookmark} 
\bookmarksetup{numbered, open,}
\usepackage{enumitem}
\setlist[enumerate]{itemsep=0mm}
\usepackage{hyperref}
\usepackage{subfigure}
\usepackage{xspace}
\usepackage{xcolor} 
\usepackage{soul} 
\usepackage{comment}
\DeclareRobustCommand{\VAN}[3]{#2}
\let\VANthebibliography\thebibliography
\def\thebibliography{\DeclareRobustCommand{\VAN}[3]{##3}\VANthebibliography}

\usepackage{orcidlink}
\usepackage{tablefootnote}
\usepackage{todo}
\usepackage[flushleft]{threeparttable}
\usepackage{ulem}
\usepackage{subfigure}
\usepackage{multirow}
\usepackage{array}
\usepackage{hyperref}

\newcolumntype{H}{>{\setbox0=\hbox\bgroup}c<{\egroup}@{}}
\newcommand{\msolar}{$M_{\odot}$}

\newcommand{\kms}{km~s$^{-1}$}

\newcommand{\Cofs}{\ensuremath{^{56}}Co\xspace}
\newcommand{\Nifs}{\ensuremath{^{56}}Ni\xspace}

\newcommand{\mic}{\ensuremath{\mu}m\xspace}

\newcommand{\acko}{SN~2022acko\xspace}
\newcommand{\jwst}{{\it JWST}\xspace}

\usepackage{newunicodechar,graphicx}
\DeclareRobustCommand{\okina}{%
  \raisebox{\dimexpr\fontcharht\font`A-\height}{%
    \scalebox{0.8}{`}%
  }%
}
\newunicodechar{ʻ}{\okina}

\graphicspath{{./}{Figures/}}

\received{XXX}
\revised{\today}
\accepted{xxx}

\submitjournal{ApJ}

\shorttitle{JWST observations of SN~2022acko}
\shortauthors{Medler, Mera, Ashall, et al.}

\begin{document}
\title{\textbf{\textit{JWST}} Medium-Resolution Infrared Spectroscopy of \acko: Tracing Molecule Formation in the Nebular Phase}

\correspondingauthor{Kyle Medler}
\email{kmedler@hawaii.edu}

\input{authors.tex}

\begin{abstract} 

The Type II supernova (SN\,II) \acko\ was the first to be spectroscopically observed by the James Webb Space Telescope (\jwst). Here, we analyze \acko’s second and third \jwst\ spectra obtained at $+259$ and $+368$~d. We identify strong features associated with hydrogen along with Intermediate-Mass and Iron-Group Elements (IM/IGEs). The medium-resolution mode of \jwst/MIRI uniquely enables the isolation of emission features, allowing us to determine the structure of \acko, directly coupling the spectroscopic features and the explosion mechanism. We find that IMEs display peak velocities of $\sim300$~\kms, significantly larger than the $\sim100$~\kms\ measured for H, He, and IGEs. We suggest a bipolar outflow best explains this ejecta distribution, although Rayleigh-Taylor instabilities may also contribute. Additionally, we find a bulk velocity offset of $\sim 97.4^{+86.3}_{-42.3}$~\kms\ in the ejecta which we associate with the natal kick of a neutron star. CO emission is also detected while no SiO or dust signatures are observed. We fit the CO first-overtone and fundamental bands with MOFAT and find a clumped distribution is required with a CO mass increasing from $1.55\times10^{-4}$~\msolar\ at $+259$ to $2.47\times10^{-4}$~\msolar\ at $+368$~d. This CO mass is approximately an order of magnitude lower than that of SN\,2024ggi. As the first \jwst\ nebular-phase study of a low-mass SN\,II, this work shows that such events form substantially less molecules than more massive SNe\,II, with dust formation likely occurring on longer timescales, if at all.

\end{abstract}


\section{Introduction} \label{sec:intro}

Type II supernovae (SNe\,II) are the explosions of H-rich massive stars with zero age main sequence masses above $\sim 8$~\msolar. Owing to their short progenitor lifetimes, they are expected to play an important role in rapidly enriching their host galaxies with heavy elements and are often invoked as a significant source of dust at early cosmic times \citep{Dwek_1980, Todini_2001, Schneider_2004, Maiolino_2004, Dwek_2006, Gall_2011_a}. A key transition in this evolution is the cooling of the ejecta from an atom-dominated plasma into the molecule formation regime. Late-time infrared spectroscopy is uniquely suited to exploring this transition because it contains several of the principal diagnostics of the cooling inner ejecta, and therefore provides direct constraints on the onset and strength of molecular emission, as well as on the cooling, mixing, and structure of the ejecta \citep{Liu_1995, Wooden1993, Cherchneff_2008, Sarangi_2013,Medler_2025,Mera_2026}.

Among the first molecules expected to form in SNe\,II are carbon monoxide (CO) and silicon monoxide (SiO). These species trace the temperature, density, and chemical stratification of the ejecta, and because they can provide direct constraints on the degree of mixing between the H-rich envelope and the carbon-oxygen rich core \citep{Liu_1992, Wooden1993, Sarangi_2013}. The principal diagnostics are the CO first overtone near 2.3~\micron, the CO fundamental band near 4.4~\micron, and the SiO fundamental band near 8.1~\micron\ \citep{Snow_1929, Singh_1975, Wooden1993}. However, historically these features have not been observed in a uniform way. Limited ground based studies have previously detected the CO first overtone in a limited number of events \citep{Catchpole_1987, Spyromilio_1988, Spyromilio_1996, Gerardy_2000, Kotak_2005, Pozzo_2007, Meikle_2011, Rho18, Davis_2019}, while \textit{Spitzer} provided only partial coverage of the CO fundamental band, although it enabled the first promising detections of SiO in SNe\,II \citep{kotak06, Kotak_2009, Szalai_2011, szalai13}. As a result, the emergence of molecular emission and its connection to the broader ejecta spectrum have remained only weakly constrained.

The advent of \jwst\ has changed this picture. With sensitive spectroscopy spanning the Near- and Mid-Infrared (NIR/MIR), \jwst\ can follow nearby SNe\,II from the plateau phase into the nebular regime and directly connect the early H dominated spectra to the subsequent evolution of the inner ejecta. Plateau phase observations of \acko\ \citep{Shahbandeh_2024}, SN\,2023ixf \citep{DerKacy_2026}, and SN\,2024ggi \citep{Baron_2025, Mera_2026} showed infrared spectra dominated by H emission. At later times, however, \jwst\ reveals a different picture. In SN\,2023ixf, strong CO emission together with an emerging infrared excess shows that \jwst\ can trace both molecular and dust emission \citep{Medler_2025}. In SN\,2024ggi, modeling of the CO bands shows that the molecular emission can be used to constrain the multidimensional structure of the ejecta \citep{Mera_2026},  demonstrating that late time infrared spectroscopy provides direct constraints on the chemistry, geometry, and cooling of SNe\,II.

SN\,2022acko is particularly important in this context. As the first SN\,II spectroscopically observed with \jwst\ \citep{Shahbandeh_2024}, \acko\ provides the first opportunity to establish a high resolution infrared baseline for a H-rich core collapse event with \jwst. \acko\ was discovered on UT 2022 December 6.2 by DLT40 in NGC~1300 and was rapidly classified as a young SN\,II \citep{DLT40, Lundquist_2022, Li22}. Throughout this work we adopt the PHANGS distance to NGC~1300 of $18.99 \pm 2.85$~Mpc \citep{2021MNRAS.501.3621A}. Pre explosion imaging suggests a red supergiant progenitor with an initial mass between $\sim 7.7$ and $9$~\msolar\ and little evidence for substantial pre-existing circumstellar dust \citep{VanDyk_2023}. Optical and ultraviolet observations further showed that \acko\ is a low luminosity SN\,IIP, making it especially interesting for testing how molecule formation proceeds in a lower-energy H-rich explosion \citep{Bostroem23}.

The first \jwst\ spectrum of \acko. Obtained at $\sim +50$~d after explosion, this observation revealed numerous H\,I features together with lines associated with CNO processed material and heavier species, but no clear signatures of CO or SiO \citep{Shahbandeh_2024, DerKacy_2026}. These data therefore established that \acko\ was molecule poor at early times and provided a critical baseline for later observations. In this paper, we present NIR/MIR nebular phase spectral observations of \acko\ obtained at $+259$ and $+368$~d past explosion. These observations probe the SN as it evolves through the nebular phase, when the NIR and MIR spectra become increasingly sensitive to the inner ejecta, nebular cooling, and molecule formation. 

Our work on the second and third \jwst\ observation of \acko\ is structured as follows. The ground-based and \jwst\ observations are described in Section~\ref{sec:obs}. We discuss the initial properties of the $1 - 16$~\micron\ spectrum of \acko\ and compare these latest observations with the initial \jwst\ observation of \acko\ in Section~\ref{sec:SED}. Next in Section~\ref{sec:IDs}, the strongest spectral features are identified and the \jwst\ observations of \acko\ are compared to other SNe\,II possessing late-time NIR and MIR spectroscopy in Section~\ref{sec:Spec_comp}. Then the velocity structure of the ejecta in \acko\ found from analyzing the emission lines is discussed in Section~\ref{sec:vel}. Section~\ref{sec:CO_form}, discusses the molecular emission and model the relevant features. Next, the analysis discussed in the previous two sections are combined and the progenitor structure required to explain the \jwst\ observations of \acko\ is presented in Section~\ref{sec:structure}. Finally, our conclusions are summarized in Section~\ref{sec:conc}.

\begin{deluxetable}{lcc}
  \tablecaption{Basic properties of \acko\ and its host galaxy \label{tab:22ackodetails}} 
  \tablehead{\colhead{Parameter} & \colhead{Value} & \colhead{Reference}}
  \startdata
  RA &03:19:38.990& (1)\\
  Dec & $-$19:23:42.68& (1)\\
  Redshift & 0.00526& (2)\\
  Host galaxy & NGC~1300& (1)\\
  $E(B-V)_{MW}$ [mag]  & $0.026 \pm 0.001$& (3)\\
  $E(B-V)_{host}$ [mag] & $0.03 \pm 0.01$& (4)\\
  Distance [Mpc]& $18.99 \pm 2.85$ & (5)\\ 
  Explosion epoch ($t_{exp}$)\tablenotemark{a} [MJD]& $59918.17\pm0.4$ & (4)\\ 
  \enddata
  \tablerefs{(1) \citet{Lundquist_2022}, (2) \citet{Li22}, (3) \citet{Schlafly_2011}, (4) \citet{Bostroem23}, (5) \citet{2021MNRAS.501.3621A}}
\end{deluxetable}


\section{Observations}\label{sec:obs}
\setcounter{footnote}{0}

Infrared observations of \acko\ were obtained at two epochs between $\sim 290$ and $400$~d after explosion, during the nebular phase. The \textit{JWST} observations were acquired through program GO-2122 (PI: Ashall) with MIRI/MRS and NIRSpec, while the ground based NIR spectra were obtained with NIRES on Keck II as part of the Hawaii Infrared Supernova Study (HISS; \citealt{Medler25}). Program GO-2122 was designed to obtain a time series of infrared spectra of a SN~II spanning the plateau phase through the onset of molecule formation. The first epoch, obtained at +50~d with NIRSpec and MIRI/MRS, was initially presented in \citet{Shahbandeh_2024} and a re-reduction presented in \citet{DerKacy_2026} which is the version used in the analysis in this work.

The reduction procedures for the individual spectra are described in the following subsections. Throughout this work, all \jwst\ data were reduced using Calibration Pipeline version~1.18.0 together with Calibration Reference Data System (CRDS) version~12.1.5 using \texttt{jwst\_1364.pmap} \citep{Bushouse_2025_JWSTpipeline}. A log of the observations is provided in Tables~\ref{tab:info_ep2} and~\ref{tab:info_ep3}.

\begin{figure*}
    \centering
    \includegraphics[width=\linewidth]{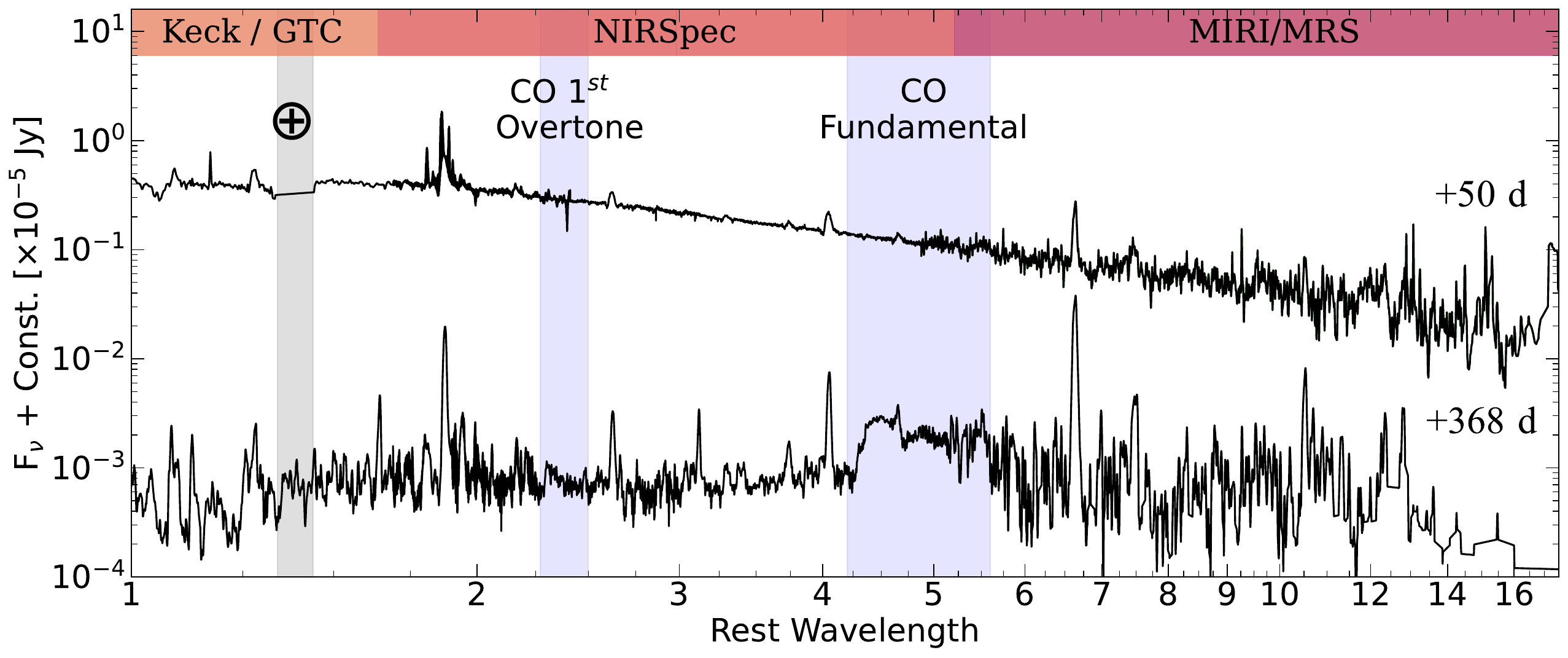}
    \caption{1-16\,\micron\ SED of \acko\ at t $\approx +50$~d and $+368$~d. The first epoch was constructed using ground-based GTC/EMIR and the \jwst\ data presented in \citep{Shahbandeh_2024,DerKacy_2026}. The SED of the third epoch was constructed using ground-based Keck/NIRES data and the \jwst\ data presented in this work. The NIRES data have been scaled to match the flux of the NIRSpec data. A smoothing Gaussian function was applied to the MIRI/MRS data with $\sigma = 3$ for $+50$~d and $\sigma = 5$ for $+368$~d spectrum. In addition, the $+368$~d SED has been scaled down by a factor of $10$ for clarity. Strong H emission lines are observed in both epochs, with emission features from IME and IGE emerging between the two epochs.}
    \label{fig:SED_1+3}
\end{figure*}

\begin{figure*}
    \centering
    \includegraphics[width=\linewidth]{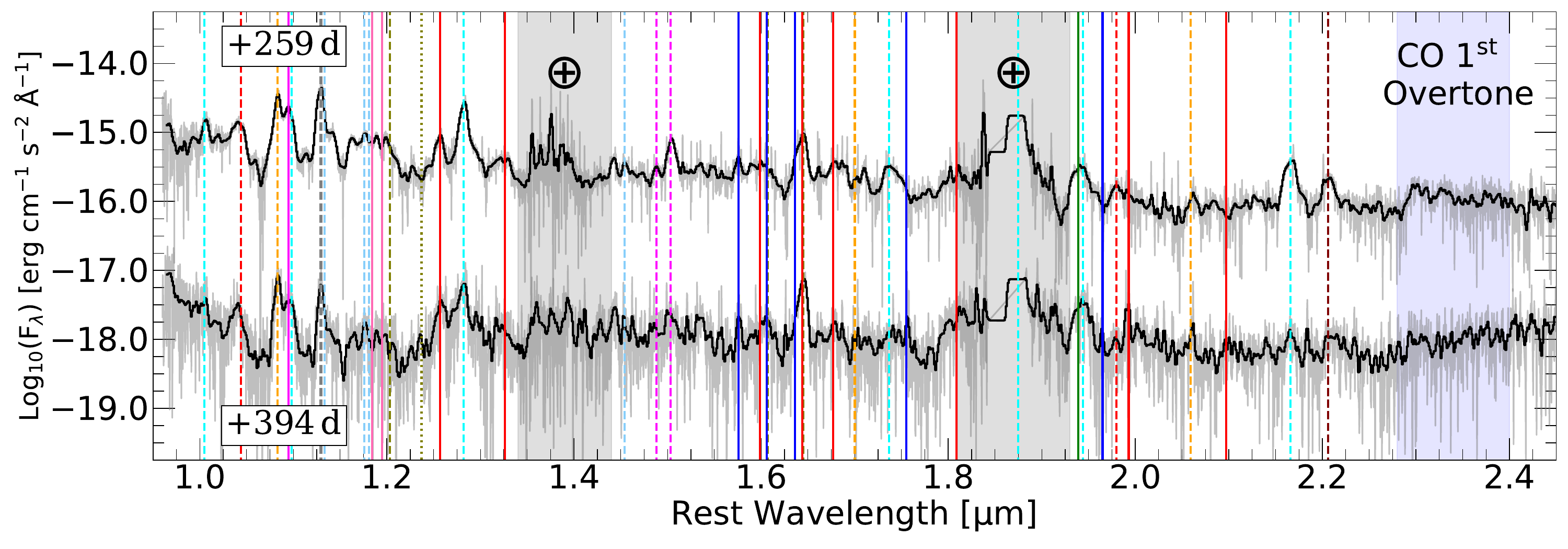}
    \includegraphics[width=\linewidth]{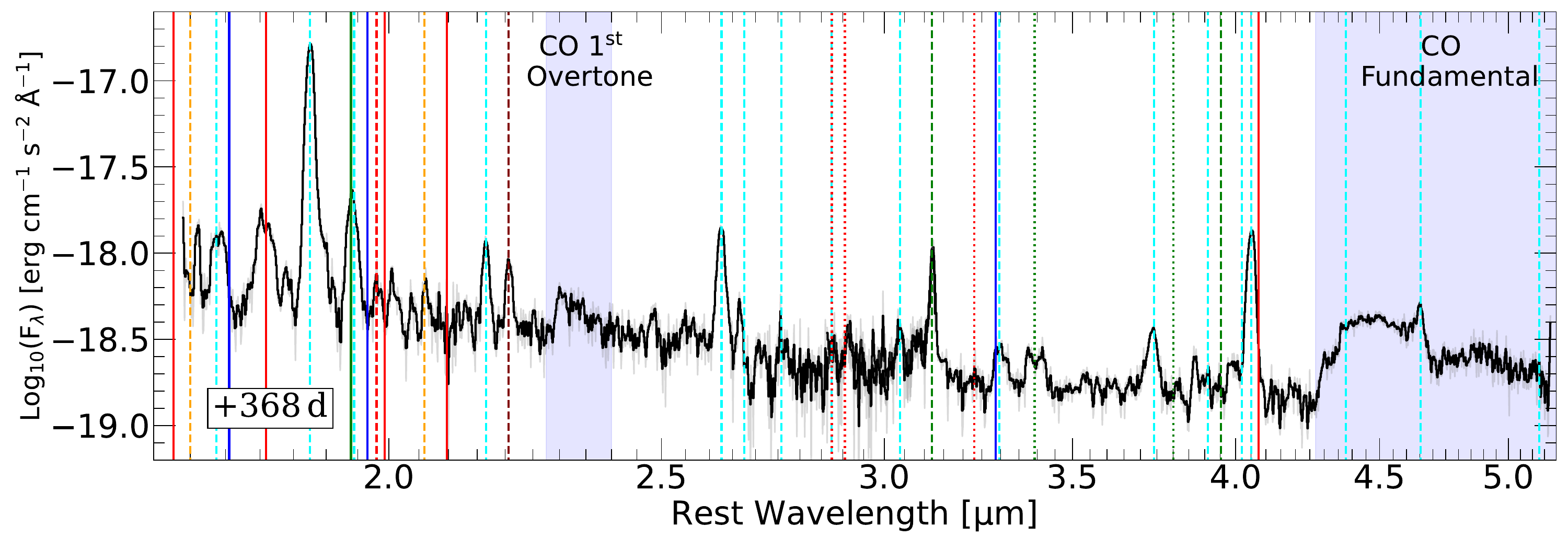}
    \includegraphics[width=\linewidth]{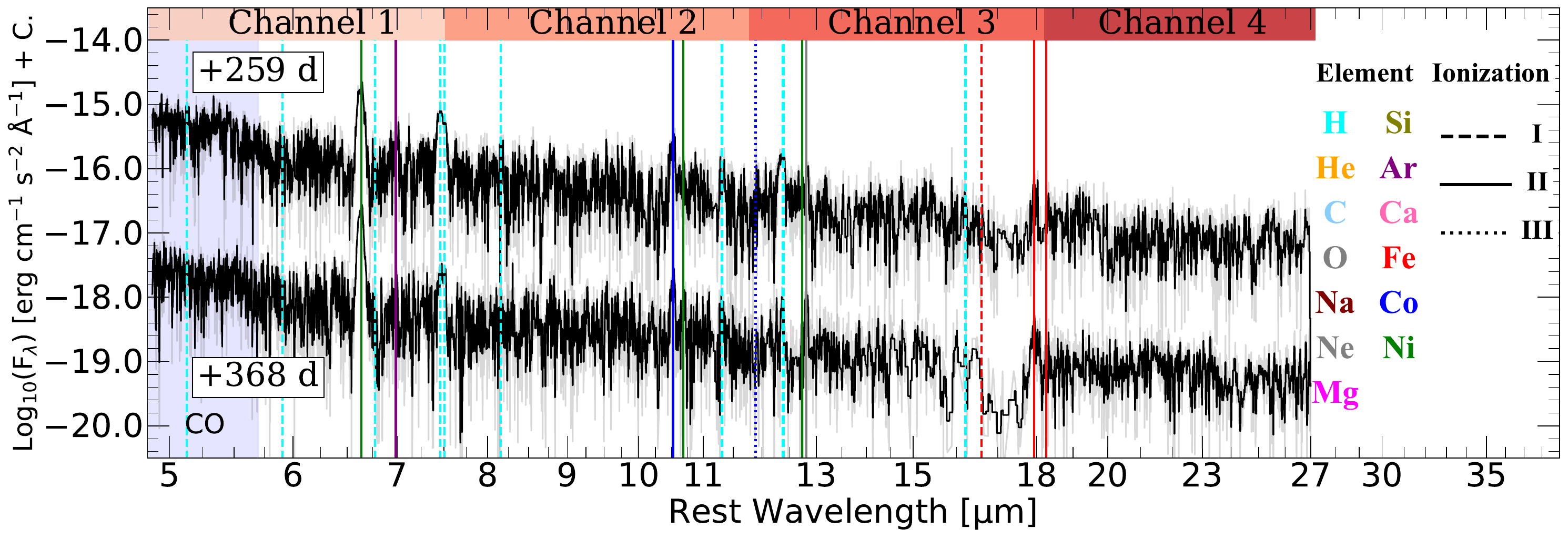}
    \caption{Line identifications for the ground-based Keck/NIRES spectra ($0.9$--$2.4$~\mic; top), \textit{JWST}/NIRSpec spectrum ($1.7$--$5.2$~\mic; middle), and \textit{JWST}/MIRI MRS spectra ($5.0$--$27.0$~\mic; bottom). All spectra have been corrected for extinction and redshift. Wavelengths and identifications are listed in Table~\ref{tab:line_IDs}. Due to known throughput issues in MIRI Channel 4, features in this region should be interpreted with caution.}
    \label{fig:Line_IDs}
\end{figure*}

\subsection{MIRI Medium Resolution Spectroscopy}
Medium-resolution spectroscopic data were obtained with MIRI at two epochs: $+259$~d and $+368$~d post-explosion. Observations were performed using the Short, Medium, and Long wavelength gratings to provide continuous spectral coverage from 5--25~\micron. A 4-point dither pattern was employed at each epoch. The data were reduced using the publicly available MIRI/MRS reduction notebook\footnote{\url{https://github.com/spacetelescope/jwst-pipeline-notebooks/blob/main/notebooks/MIRI/MRS/JWPipeNB-MIRI-MRS.ipynb}}, in order to produce one continuous data cube comprising all channels and sub-bands. 

The background flux in the MIRI/MRS data is poorly characterized and varies both spatially across the field of view and as a function of wavelength \citep[see e.g.,][]{Shahbandeh_2024, DerKacyxkq, Ashall24}. To remove this background, comprising instrumental and telescope noise as well as complex emission from the host galaxy, we employ the custom-built background subtraction pipeline AstroBkgInterp\footnote{\url{https://github.com/brynickson/AstroBkgInterp}} \citep{AstroBkg}.  In brief, the pipeline estimates the two-dimensional background beneath the SN at each wavelength using a combination of interpolation and polynomial surface fitting. The code requires aperture extraction radius and an outer annulus width, which we set to 10 and 3 pixels, respectively. We tested larger aperture sizes for the extraction radius and found that it produced more noise in the final extracted spectrum without changing the line fluxes.  The flux of the SN is then extracted from the full data cube the \textsc{Extract1dStep} in the \textsc{JWST} pipeline. 

We note that the data redward of $\sim$20~$\micron$ are dominated by noise and instrumental background. This is in part because \acko\ was a low luminosity SN\,II, but is enhanced by known issues with the degrading count rate throughput of Ch3 and Ch4 \footnote{\url{https://www.stsci.edu/contents/news/jwst/2023/miri-mrs-reduced-count-rate-update}}. Regardless, the SN is visible in some slices of the data cube at these longest wavelengths, and while the continuum is not accurate in these regions, tentative line IDs are reliable. Future analysis, comparing \textit{JWST} MIRI photometric observations of \acko\ \citep{2023jwst.prop.3295S} at these epochs will be able to assist in further refining the MRS flux calibration as well as determine if any weak dust emission is visible in the ejecta. To aid in data visualization we correct the channel 4 emission by fitting the MIRI/MRS continuum across Channels 1 and 2, between $5 - 10$~\micron, and Channel 4, over $17 - 27$~\micron, with a linear function. We then extrapolate the Channel 1 and 2 continuum out to Channel 4 and suppress the observed flux to match the extrapolated continuum. 

\subsection{NIRSpec}
A single NIRSpec spectrum of \acko\ was obtained at +368~d, covering $\sim 1.7$ to 5.2~\micron\ at $R \sim 1000$. The observations were taken with the S400A1 slit using the G235M/F170LP and G395M/F290LP configurations. Two integrations per exposure and one exposure per dither was used in each grating, corresponding to a total integration time of 383.39~s per grating. The data were acquired on MJD~60287.77 using a three point dither pattern. The background subtracted level 2 products were combined to produce a single calibrated spectrum for each grating. The spectral reduction was performed with the publicly available NIRSpec pipeline notebook\footnote{https://github.com/spacetelescope/jwst-pipeline-notebooks/blob/main/notebooks/NIRSPEC/FSlit/JWPipeNB-NIRSpec-FS.ipynb}. Details of the observational setup are given in Table~\ref{tab:info_ep3}.

\subsection{Ground-based spectroscopy}
To complete the Spectral Energy Distribution (SED) at shorter wavelengths, we obtained ground-based NIR spectra of \acko\ at similar epochs through HISS \citep{Medler25}. The ground-based NIR observations were obtained using the Near-Infrared Echellette Spectrometer (NIRES; \citealt{Adkins2014}) mounted on the Keck-II telescope. Data were acquired using an ABBA dither pattern, with individual exposure times of 300~s. The data were reduced using the IDL package \textsc{Spextool} version~5.0.2 \citep{Cushing_2004_Spextool}. Flux calibration and telluric correction were performed using \textsc{Xtellcorr} version~5.0.2, based on contemporaneous observations of A0V standard stars.
We also show a new ground-based NIR spectrum of \acko\ obtained by the Espectrógrafo Multiobjeto Infra-Rojo (EMIR) mounted on the 10.4-m Gran Telescopio Canarias (GTC). This data was taken on Jan. 07 2023, at $+33.6$~d, close to the initial \jwst\ observation. NIR observations were obtained at three epochs with the $YJ$ and $HK$ grisms. The data were reduced using custom pipelines following \cite{2025JCAP...08..053G}. The EMIR reduction incorporates routines from \texttt{pyemir} \citep{2010ASPC..434..353P}. Standard reduction procedures were applied, including detector-level calibrations, wavelength calibration, sky subtraction, and one-dimensional spectral extraction. Relative flux calibration was performed using spectrophotometric standard stars observed immediately following the science observations for the NIR data. Individual exposures were combined to produce the final calibrated spectra used in this work.
A log of the ground-based observations is given in Table~\ref{tab:GB_info}.

 \section{Spectral Energy Distribution Evolution} \label{sec:SED}
 In Figure~\ref{fig:SED_1+3} we show the ground-based NIR and \jwst\ NIRSpec and MIRI/MRS spectra of \acko\ and compare the SED from $+368$~d to the SED obtained at $+50$~d from \citet{Shahbandeh_2024, DerKacy_2026}. Between the two epochs there is a clear decline in the SED of \acko\ as the SN cools and fades. Unlike other SNe\,II observed at MIR wavelengths, the IR continuum of \acko\ shows no signs of an infrared excess associated with dust \citep{Kotak_2009, Szalai_2011, Shahbandeh_2023, Medler_2025}. Figure~\ref{fig:SED_1+3} highlights the main differences between the initial epoch of \acko\ and the latest, with the strong H emission features and the [\ion{Ni}{2}]~$6.636$~\micron\ being observed in both epochs. However the two SEDs differ significantly at other wavelengths showing the emergence of features associated with strong Intermediate-Mass and Iron-Group Elements (IM/IGEs) lines, along with CO emission features, which are present at the later phase. We discuss the lines identified in the section below. 

\begin{table*}
\centering
\small
\setlength{\tabcolsep}{4pt}
\caption{Line identifications in the spectra of \acko}
\label{tab:line_IDs}
\begin{tabular}{lclclclc}
\hline
Ion & Wavelength [$\mu$m$]$ & Ion & Wavelength [$\mu$m$]$ & Ion & Wavelength [$\mu$m$]$ & Ion & Wavelength [$\mu$m] \\
\hline

\multicolumn{8}{c}{Keck/NIRES spectrum, 0.9 to 1.7~$\mu$m} \\
\hline
Pa$\delta$ & 1.005 & \ion{Fe}{1}& 1.044 & \ion{Mg}{1} & 1.081 & \ion{Mg}{2} & 1.095 \\
Pa$\gamma$ & 1.098 & \ion{O}{1} & 1.129 & \ion{O}{1} & 1.130 & \ion{C}{1} & 1.133 \\
\ion{C}{1} & 1.176 & \ion{C}{1} & 1.176 & \ion{C}{1} & 1.181 & \ion{Ca}{2} & 1.184 \\
\ion{Ca}{2} & 1.195 & \ion{Si}{1} & 1.203 & \ion{Si}{3} & 1.237 & [\ion{Fe}{2}] & 1.257 \\
Pa$\beta$ & 1.282 & \ion{Fe}{1}& 1.326 & \ion{C}{1} & 1.454 & \ion{Mg}{1} & 1.488 \\
\ion{Mg}{1} & 1.503 & [\ion{Co}{2}] & 1.546 & [\ion{Fe}{2}] & 1.599 & \ion{Co}{2} & 1.606 \\
$\mathrm{[Si I]}$ & 1.607 & \ion{Co}{2} & 1.636 & [\ion{Fe}{2}] & 1.644 & \ion{Si}{1} & 1.645 \\
$[$\ion{Fe}{2}$]$ & 1.677 &  &  &  &  &  &  \\
\hline

\multicolumn{8}{c}{\textit{JWST}/NIRSpec spectrum, 1.7 to 5.2~$\mu$m} \\
\hline
Br$_{10\rightarrow4}$ & 1.737 & \ion{Co}{2} & 1.755 & [\ion{Fe}{2}] & 1.809 & Br$\varepsilon$ & 1.817 \\
Pa$\alpha$ & 1.875 & \ion{Co}{2} & 1.914 & [\ion{Ni}{2}] & 1.939 & Br$\delta$ & 1.944 \\
$[$\ion{Co}{2}$]$ & 1.965 & \ion{Fe}{1}& 1.980 & \ion{Fe}{2}& 1.993 & \ion{Fe}{2}& 2.061 \\
$[$\ion{Co}{2}$]$ & 2.097 & Br$\gamma$ & 2.166 & \ion{Na}{1} & 2.206 & Br$\beta$ & 2.626 \\
Pf$_{13\rightarrow5}$ & 2.675 & Pf$_{12\rightarrow5}$ & 2.758 & Pf$_{11\rightarrow5}$ & 2.873 & [\ion{Fe}{3}] & 2.874 \\
$[$\ion{Fe}{3}$]$ & 2.905 & Pf$\varepsilon$ & 3.039 & [\ion{Ni}{1}] & 3.120 & [\ion{Fe}{3}] & 3.229 \\
$[$\ion{Co}{2}$]$ & 3.286 & Pf$\delta$ & 3.297 & [\ion{Ni}{3}] & 3.394 & Pf$\gamma$ & 3.741 \\
$[$\ion{Ni}{3}$]$ & 3.802 &  Hu$_{15\rightarrow6}$ & 3.910 & [\ion{Ni}{1}] & 3.952 & Hu$_{14\rightarrow6}$ & 4.02 \\
Br$\alpha$ & 4.051 & [\ion{Fe}{2}] & 4.076 & Hu$_{12\rightarrow6}$ & 4.377 & Pf$\beta$ & 4.654  \\
Hu$\delta$ & 5.129 &  &  \\
\hline

\multicolumn{8}{c}{\textit{JWST}/MIRI MRS spectrum, 5.0 to 28~$\mu$m} \\
\hline
Hu$\gamma$ & 5.908 & [\ion{Ni}{2}] & 6.636 & H$_{12\rightarrow7}$ & 6.772 & [\ion{Ar}{2}] & 6.985 \\
Pf$\alpha$ & 7.460 & Hu$\beta$ & 7.503 & H$_{15\rightarrow8}$ & 8.155 & H$_{14\rightarrow8}$ & 8.664 \\
$[$\ion{Co}{2}$]$ & 10.521 & [\ion{Ni}{2}] & 10.682 & [\ion{Ni}{1}] & 11.307 & H$_{9\rightarrow7}$ & 11.309 \\
$[$\ion{Co}{3}$]$ & 11.886 & [\ion{Co}{1}] & 12.255 & Hu$\alpha$ & 12.372 & H$_{11\rightarrow8}$ & 12.390 \\
$[$\ion{Ni}{2}$]$ & 12.729 & [\ion{Ne}{2}] & 12.813 & [\ion{Co}{3}] & 13.813 & H$_{17\rightarrow10}$ & 13.942 \\
$[$\ion{Co}{3}$]$ & 13.996 & H$_{13\rightarrow9}$ & 14.183 & H$_{16\rightarrow10}$ & 14.962 & [\ion{Fe}{1}] & 15.561 \\
$[$\ion{Fe}{1}$]$ & 16.107 & H$_{10\rightarrow8}$ & 16.209 & [\ion{Fe}{1}] & 16.596 & H$_{12\rightarrow9}$ & 16.881 \\
$[$\ion{Fe}{2}$]$ & 17.936 & [\ion{Ni}{2}] & 18.240 & [\ion{Fe}{2}] & 18.267 & H$_{14\rightarrow10}$ & 18.615\\
H$_{20\rightarrow12}$ & 20.515 & H$_{16\rightarrow11}$ & 20.921 & H$_{19\rightarrow12}$ & 21.842 \\
\hline
\end{tabular}
\end{table*}

\section{Spectral Emission Features} \label{sec:IDs}

The spectra of \acko\ were obtained several hundred days after explosion when the ejecta had become optically thin, it was primarily powered by the radioactive decay of \Cofs\ and the emission was dominated by forbidden line transitions. At these epochs they display prominent emission features from H, He, IMEs, IGEs, and CO molecules. 

In this work line identifications were guided by previous studies of late-time SN~II spectra \citep{Jerkstrand_2012, Davis_2019, Shahbandeh_2022, Shahbandeh_2024, Medler_2025}. The spectra separated by wavelength region are shown in Figure~\ref{fig:Line_IDs}, and the identified spectral lines are listed in Table~\ref{tab:line_IDs}. 

\subsection{Hydrogen}

H emission remains prominent throughout the late time NIR and MIR spectra of \acko. Across the ground based and \textit{JWST} observations, we identify lines from the Paschen, Brackett, Pfund, and Humphreys series. In the NIR, the strongest Paschen features include Pa$\alpha~1.875$~\micron, Pa$\beta~1.282$~\micron, and Pa$\gamma~1.094$~\micron, while the Brackett series is represented by Br$\alpha~4.051$~\micron, Br$\beta~2.625$~\micron, Br$\gamma~2.166$~\micron, Br$\delta~1.944$~\micron, and Br$\eta~1.817$~\micron. The \textit{JWST} spectra also reveal a rich set of Pfund lines, including Pf$\alpha~7.466$, Pf$\beta~4.654$, Pf$\gamma~3.741$, Pf$\delta~3.297$, Pf$\varepsilon~3.039$, Pf$\zeta~2.873$, and Pf$\eta~2.758$~\micron. At longer wavelengths, the MIRI/MRS data further show Humphreys emission, including Hu$\alpha~12.372$~\micron, Hu$\beta~7.503$~\micron, Hu$\gamma~5.908$~\micron, and Hu$\delta~5.129$~\micron, together with additional higher order H transitions such as H$_{9 \rightarrow 7}~11.309$~\micron\ and H$_{14 \rightarrow 10}~18.615$~\micron.

\subsection{Helium}

Two emission features located at $\sim 1$~\micron\ and $\sim 2$~\micron\ are identified in the SED of \acko\ corresponding to the emissions from the strong 
\ion{He}{1} $1.083$~\micron\ and $2.059$~\micron\ lines. Additionally, there is a weaker emission feature located around $1.7$~\micron\ which may be associated with \ion{He}{1} $1.700$~\micron\ line, however this line is much weaker than the $1$ and $2$~\micron\ lines and is only identifiable in the later epochs.  

\subsection{Intermediate-Mass Elements}

Several intermediate mass elements are detected in the late time spectra of \acko. In the NIR, we identify emission from C, O, Mg, Si, and Ca, including [\ion{C}{1}]~$1.133$~\micron, [\ion{O}{1}]~$1.129$ and $1.130$~\micron, [\ion{Mg}{1}]~$1.488$ and $1.503$~\micron, [\ion{Si}{1}]~$1.607$~\micron, and [\ion{Ca}{2}]~$1.184$~\micron. These transitions trace chemically processed material in the inner ejecta and indicate that the spectra are probing relatively cool, low ionization regions. At longer wavelengths, the MIR spectra reveal additional ionized species, most notably [\ion{Ar}{2}]~$6.985$~\micron\ and [\ion{Ar}{2}]~$12.813$~\micron. The presence of both neutral and singly ionized species points to a structured ejecta with distinct emitting zones and varying ionization and excitation states.

\subsection{Iron-Group Elements}

Emission from IGEs is also prominent throughout the late time spectra of \acko, showing that the observations probe deep, chemically processed regions of the ejecta. Emission from Fe and Co is seen in several ionization states, including [\ion{Fe}{2}]~1.644~\micron, [\ion{Fe}{3}]~$2.905$ and $3.229$~\micron, [\ion{Fe}{1}]~$15.561$ and $16.596$~\micron, [\ion{Fe}{2}]~$17.936$ and $18.267$~\micron, [\ion{Co}{2}]~$1.636$ and $10.521$~\micron, [Co~I]~$12.255$~\micron, and [\ion{Co}{3}]~$13.813$~\micron. An additional feature near $\sim 4.3$~\micron\ may also be associated with a weak Fe transition, although this remains uncertain. Ni is likewise present, with [\ion{Ni}{1}]~$3.120$ and $3.952$~\micron\ and [\ion{Ni}{2}]~$6.636$, $10.682$, $12.729$, and $18.240$~\micron. The presence of Fe, Co, and Ni in multiple ionization states points to a structured nebular core with a range of excitation and ionization conditions.

\subsection{CO  Emission}
Molecular emission is clearly present in the late time spectra of \acko. The \textit{JWST}/NIRSpec spectrum reveals both the CO first overtone band at $2.29$ to $2.4$~\micron\ ($\Delta \nu = 2 \rightarrow 0$) and the CO fundamental band at $4.27$ to $5.8$~\micron\ ($\Delta \nu = 1 \rightarrow 0$). The fundamental band is substantially stronger than the first overtone, consistent with previous observations of SNe\,II \citep{Wooden1993, Medler_2025, Mera_2026}. Both CO features are nevertheless weak relative to the surrounding continuum and nearby H lines, with the first overtone peaking at $\sim 0.4 \times 10^{-5}$~Jy above the continuum and the fundamental reaching $\sim 2.0 \times 10^{-5}$~Jy. The red end of the CO fundamental band extends beyond the NIRSpec wavelength range, and its blue edge is also visible at the short wavelength end of the MIRI/MRS spectrum, providing additional support for the identification. The detection of both bands shows that molecule formation has occurred in the ejecta of \acko. At the same time, their relatively low fluxes suggest that the molecular component remains modest at these epochs. Because \acko\ was a low-luminosity SN\,II, these observations provide a useful view of how CO formation proceeds in a lower energy H-rich explosion. A quantitative analysis of the CO emission and the implied molecular yield is presented in Section~\ref{sec:CO_form}.

We searched the two MIRI spectra for evidence of SiO emission, particularly the fundamental vibrational band near $8.1$~\micron\ ($\Delta \nu = 0 \rightarrow 1$). However, no clear SiO emission is detected at either $+259$~d or $+368$~d spectra, even after continuum subtraction and significant smoothing. Likewise, neither of the MIRI/MRS spectra of \acko\ show any strong infrared continuum excess that would indicate dust emission at these epochs, unlike other H-rich SNe \citep{Kotak_2009, Medler_2025}.

\section{Spectral Comparison} \label{sec:Spec_comp}

A comparison of \acko's NIR and MIR spectra with observations of other SNe\,II allow us to better understand how this low-luminosity object differs from more typical SNe\,II. This comparison allows us to assess differences in the characteristics of CO signatures and the underlying continuum, which may indicate the presence of warm dust within the system.

\subsection{NIR Comparison}\label{sec:NIR_comp}
\begin{figure}
    \centering
    \includegraphics[width=\columnwidth]{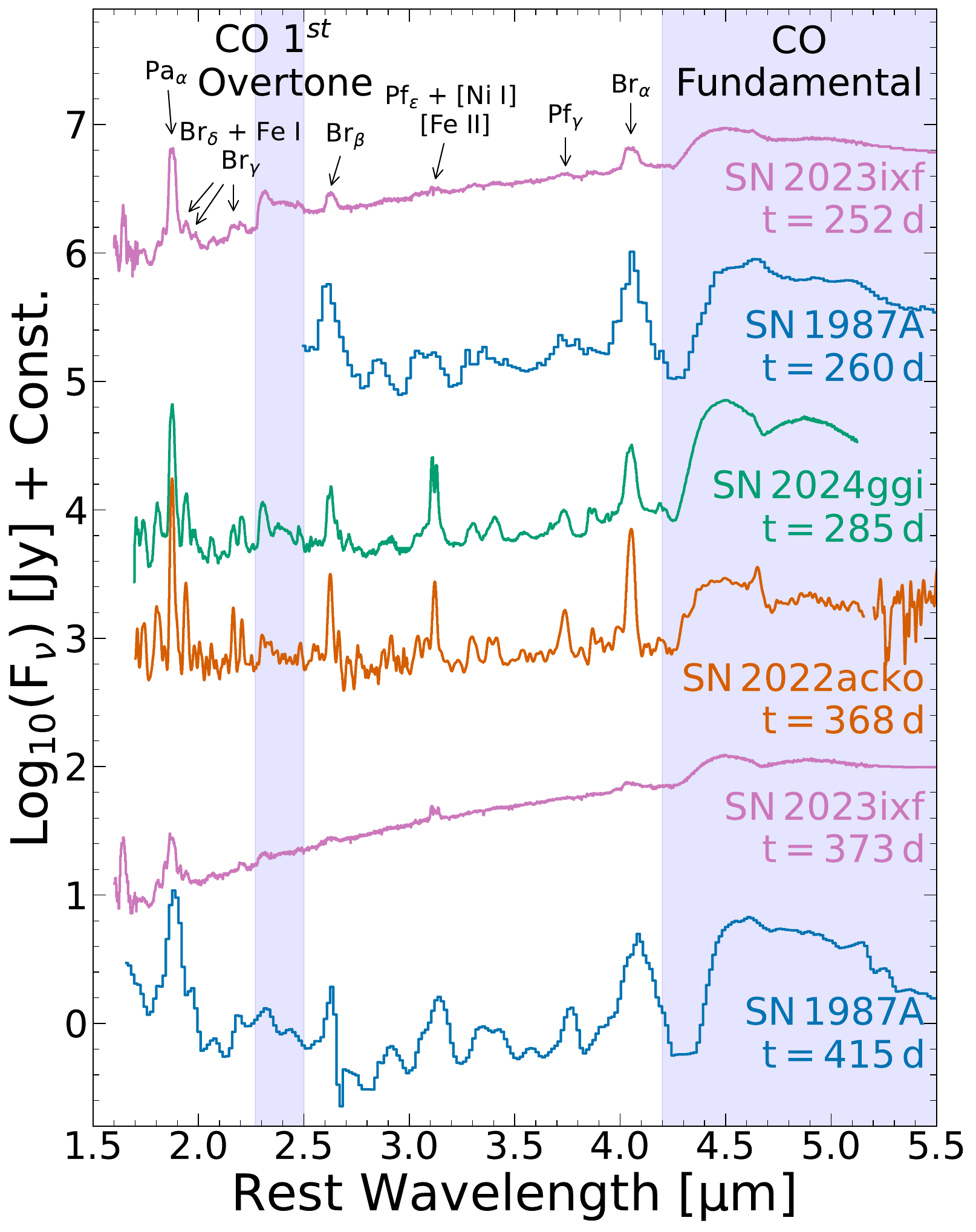}
    \caption{Comparison of the \jwst\ NIRSpec observation of \acko\ at day $+368$ with the late time observations of SN\,1987A obtained between $+250$ and  $+420$~d \citep{Wooden_1993}, as well as SN\,2023ixf \citep{Medler_2025, DerKacy_2026} and SN\,2024ggi \citep{Dessart2025, Baron_2025}. All spectra have been corrected for extinction and redshift. The CO fundamental of \acko\ is much narrower than the features observed in SN\,1987A. Additionally, the H features of \acko\ are also sharper than any other SN. The strong emission features common among the different SNe\,II are labeled.}
    \label{fig:NIR_comp}
\end{figure}

Historically, the lack of spectral coverage between $2.3$--$5.2$~\micron\ has limited comprehensive studies of simultaneous observations of the CO first overtone and fundamental bands in SNe. 
Until \acko,  SN\,1987A  was the only SN\,II with published observations fully covering both CO regions \citep{Wooden1993}. 
The NIR spectra of \acko\ along with those of SN\,1987A, SN\,2023ixf, and SN\,2024ggi observed at similar epochs are shown in Figure~\ref{fig:NIR_comp}.
Out of the three other H-rich CC-SNe compared in this work, overall \acko\ is most similar to SN\,2024ggi, with SN\,2023ixf displaying a strong dust continuum and SN\,1987A possessing broader emission profiles than either \acko\ or SN\,2024ggi.

While each SN possesses a unique infrared continuum and differing spectral emission feature properties, molecular CO emission is ubiquitous. The first CO feature observed in the NIR spectra is the CO $\mathrm{1^{st}}$ overtone which in all SNe displays a rapid rise to initial peak, followed by a broad decline down to a secondary peak near $\sim2.47$~\micron. 
Notably, a small notch on the blue side of the CO overtone can be observed in the later spectrum of SN\,1987A. This feature is produced by CO$^+$, resulting from the mixing of \Nifs\ in the ejecta. This feature is absent in \acko, as well as SN\,2023ixf and SN\,2024ggi, implying a complete lack of \Nifs\ mixing in the low-luminosity and normal SNe\,II \citep{Spyromilio_1988}.
The CO fundamental band also shows distinct differences. The initial peak of the CO fundamental is coincident with the emission feature from the Pf$_\beta~4.654$~\micron\ line. In \acko\ the Pf$_\beta$ line is significantly stronger than in either SN\,1987A, SN\,2023ixf, and SN\,2024ggi. The differences in the observed Pf$_\beta$ emission strength relative to the underlying CO feature may reflect the diversity in the H / CO mass ratio in the different SNe resulting in the varying visibility of the H line. Additionally, the Hu$_\delta~5.129$~\micron\ is more pronounced in SN\,1987A, possibly reflecting differences in CO mass that modulate the visibility of the H line. After the initial hump in the CO fundamental a smooth decline before a secondary bump is observed. The deeper drop observed between the initial peak of the CO fundamental and individual rovibrational lines in \acko\ compared to SN\,1987A and SN\,2023ixf may suggest a lower CO temperature in \acko, as discussed in Section~7.

\begin{figure*}
    \centering
    \includegraphics[width=\linewidth]{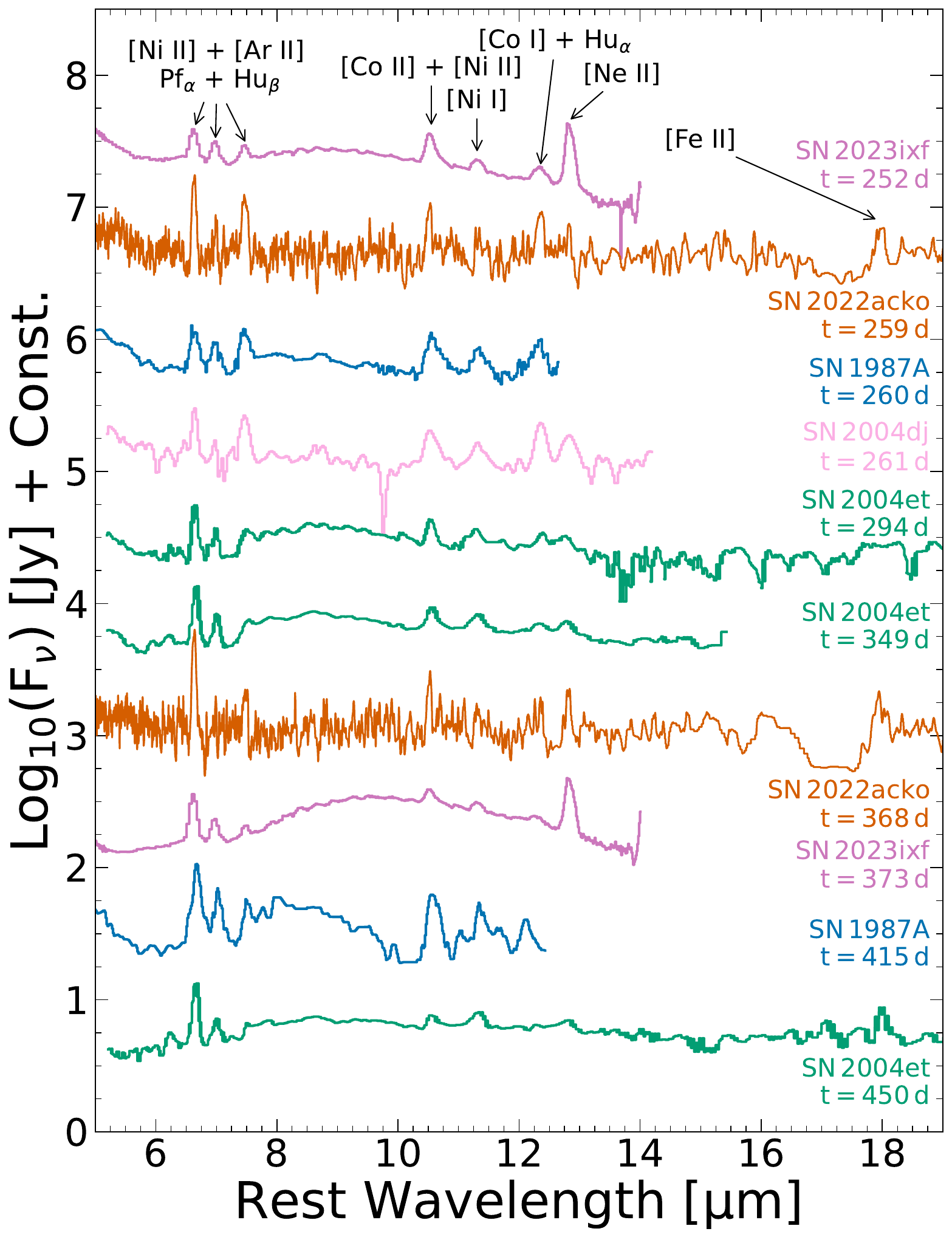}
    \caption{A comparison of several SN\,II between $5 - 18$~\mic\ between $+260 - +450$ post-explosion. Spectra taken from SN\,1987A \citep{Wooden1993}, SN\,2004dj \citep{Kotak_2005, Szalai_2011}, SN\,2004et \citep{Kotak_2009}, and SN\,2023ixf \citep{Medler_2025}.}
    \label{fig:MIR_comp}
\end{figure*}

\subsection{MIR comparison} \label{sec:MIR_comp}

We compare the MIR spectra of \acko\ with those of several other SNe\,II observed at late times, see Figure~\ref{fig:MIR_comp}. \acko, like other SNe\,II, exhibits strong H emissions along with several IGEs with intensities well above the continuum. Notably, in \acko\ the [\ion{Ni}{2}]~$6.636$~\micron\ line dominates over the adjacent [\ion{Ar}{2}]~$6.985$~\micron\ line especially at $+368$~d where the [\ion{Ar}{2}] line is just above the continuum level. In contrast the Ni/Ar emission ratio in the other SNe is much smaller, indicating that the deeper region of the ejecta was exposed significantly faster in \acko\ than in other H-rich SNe. However, it might also indicate that \acko\ produced significantly more stable Ni than Ar resulting in a stronger emission feature. 
The feature near 10.5~\micron, typically a blend of [\ion{Co}{2}]~10.52~\micron\ and [\ion{Ni}{2}]~10.682~\micron, is resolved in the higher-resolution \textit{JWST} MIRI/MRS spectra of \acko, unlike other lower resolution SNe spectra. Similarly, the $12.3$~\micron\ region, which includes a blend of [Co~I]~$12.255$~\micron\ and Hu$\alpha\,12.372$~\micron, can be fully disentangled, revealing the H feature is stronger than the Co emission. 
Finally, there is an emission feature located at $\sim 18$~\micron\ which is identified as the [\ion{Fe}{2}]~$17.936$~\micron\ line. This emission line is only observed in SN\,2004et due to the shorter wavelength coverage of the other spectra. While SN\,2004et shows a strong [\ion{Fe}{2}]~$17.936$~\micron\ at $+450$~d the earlier spectrum at $+259$~d shows a broad emission feature, possibly explained by a blend of the Fe line with the H$_{18~\rightarrow~11}$ which has faded significantly between the two epochs. The emission features in \acko\ are narrower than the same features in the MIR spectra of other SNe by several thousand \kms, a result of \acko\ possessing a lower mass and kinetic energy than typical \citep{Bostroem23}.

Another key distinction between \acko\ and other SNe\,II lies in the shape of the infrared continuum. \acko\ lacks any infrared excess between $\sim7.5$--$9.5$~\micron\ typically associated with the SiO emission region and silicate-rich dust formation \citep{Kotak_2009}. Instead, \acko\ shows a flat continuum at longer wavelengths, in stark contrast to the rising infrared continua of other SNe\,II. This absence of a strong infrared excess indicates a minimal presence of pre-existing dust around \acko\ capable of surviving the explosion. This conclusion aligns with the progenitor analysis by \cite{VanDyk_2023}, which identified a low-mass RSG as the preferred progenitor, and with early-time observations of \acko\ \citep{Shahbandeh_2024}, which also found no evidence for significant pre-existing dust. Furthermore, the lack of a late-time infrared excess strongly suggests that little to no new dust was synthesized during the evolution of \acko.

\begin{figure*}
    \centering
    \includegraphics[width=\linewidth]{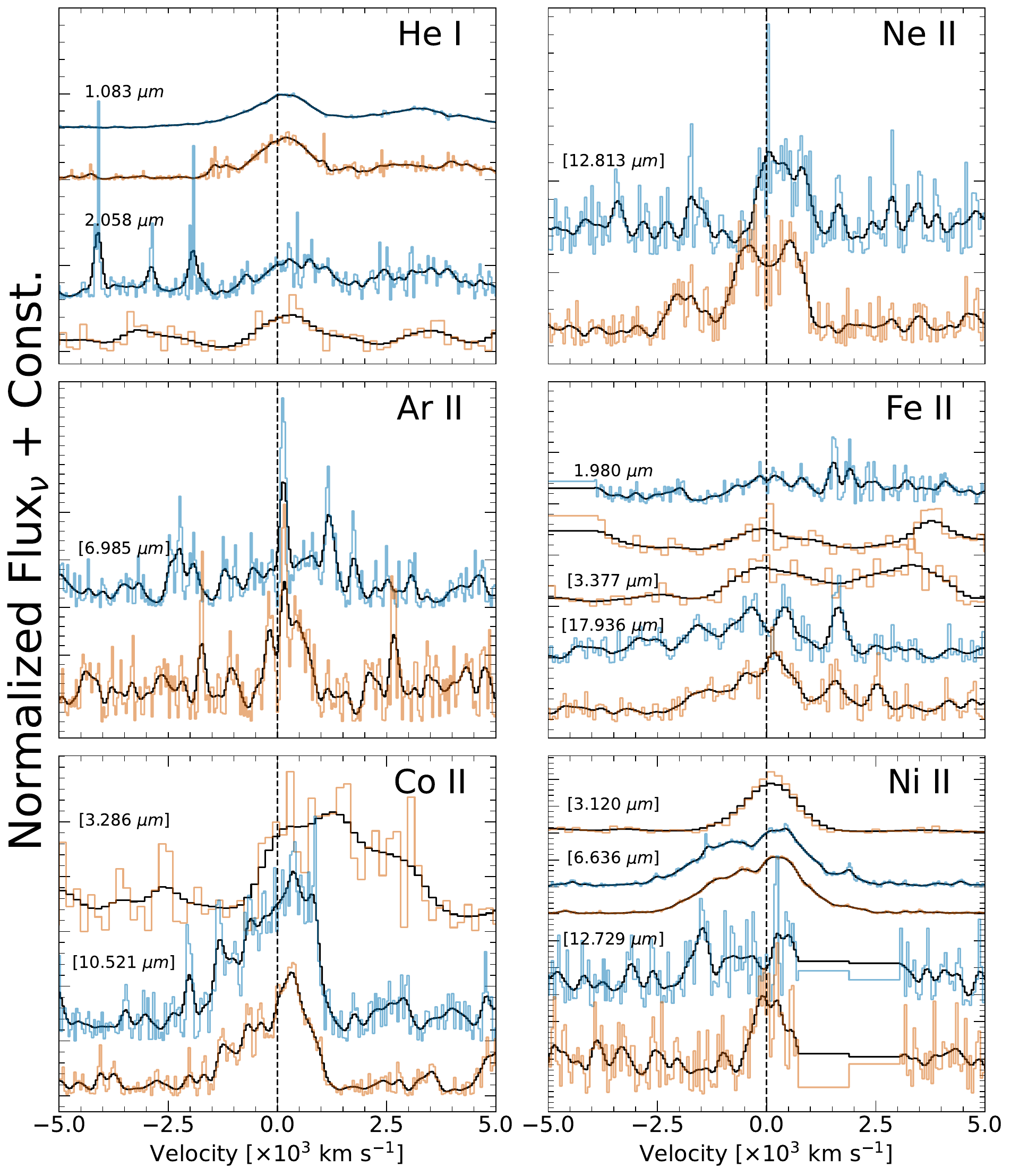}
    \caption{Comparison of the strong NIR and MIR emission features from He, Ne, Ar, Fe, Co, and Ni of \acko\ between days $+259$ (blue) and $+368$ (orange). All the MIR features shown are present at both epochs with the exception of the [\ion{Fe}{2}]~$17.936$~\mic\ which emerges within the spectrum between the observed epochs. There is a consistent shift in the peak of the emission feature by $\sim 500$~\kms. Regions effected by known emission lines are masked out.}
    \label{fig:line_vels}
\end{figure*}

\begin{figure}
    \centering
    \includegraphics[width=\linewidth]{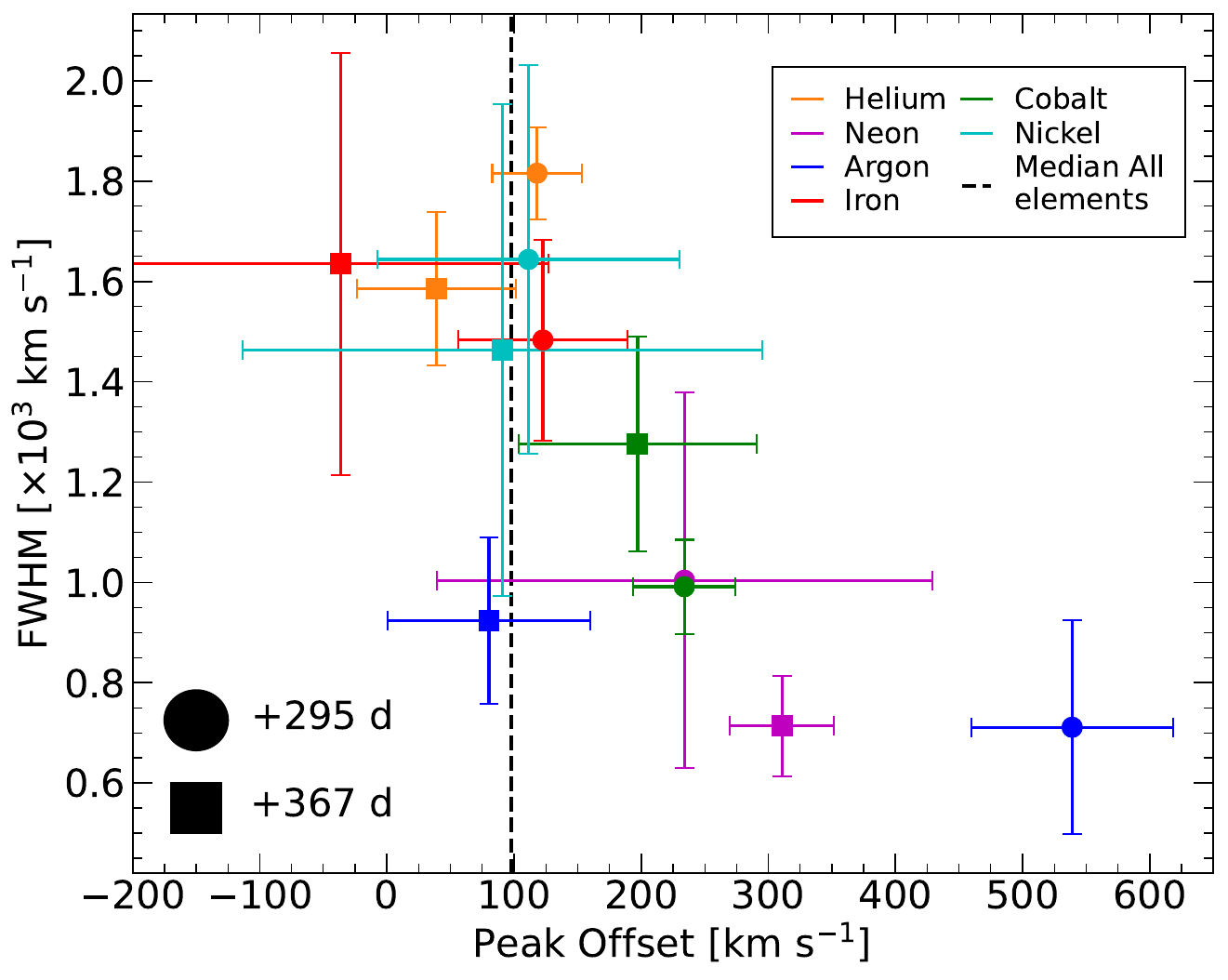}
    \caption{Comparison of the FWHM and emission peak offset for the strong NIR and MIR emission features from He, Ne, Ar, Fe, Co and Ni shown in Figure~\ref{fig:line_vels}. All the peak offsets have had the peculiar velocity of \acko\ within NGC 1300, $179.2$~\kms, subtracted. The properties determined from the $+295$~d observation are shown as circles, while the $+368$~d spectrum is shown by the square markers. There is a clear separation between the burning products (Fe, Co, and Ni) and the stellar products (Ar and Ne) suggesting strong asymmetry in the explosion, with the He shell laying far enough above the explosion to not be affected. The median offset velocity is shown by the black dashed line located at $+97.9$~\kms\ from rest.}
    \label{fig:Elem_FWHM_vs_Peak}
\end{figure}

\begin{figure*}
    \centering
    \includegraphics[width=\linewidth]{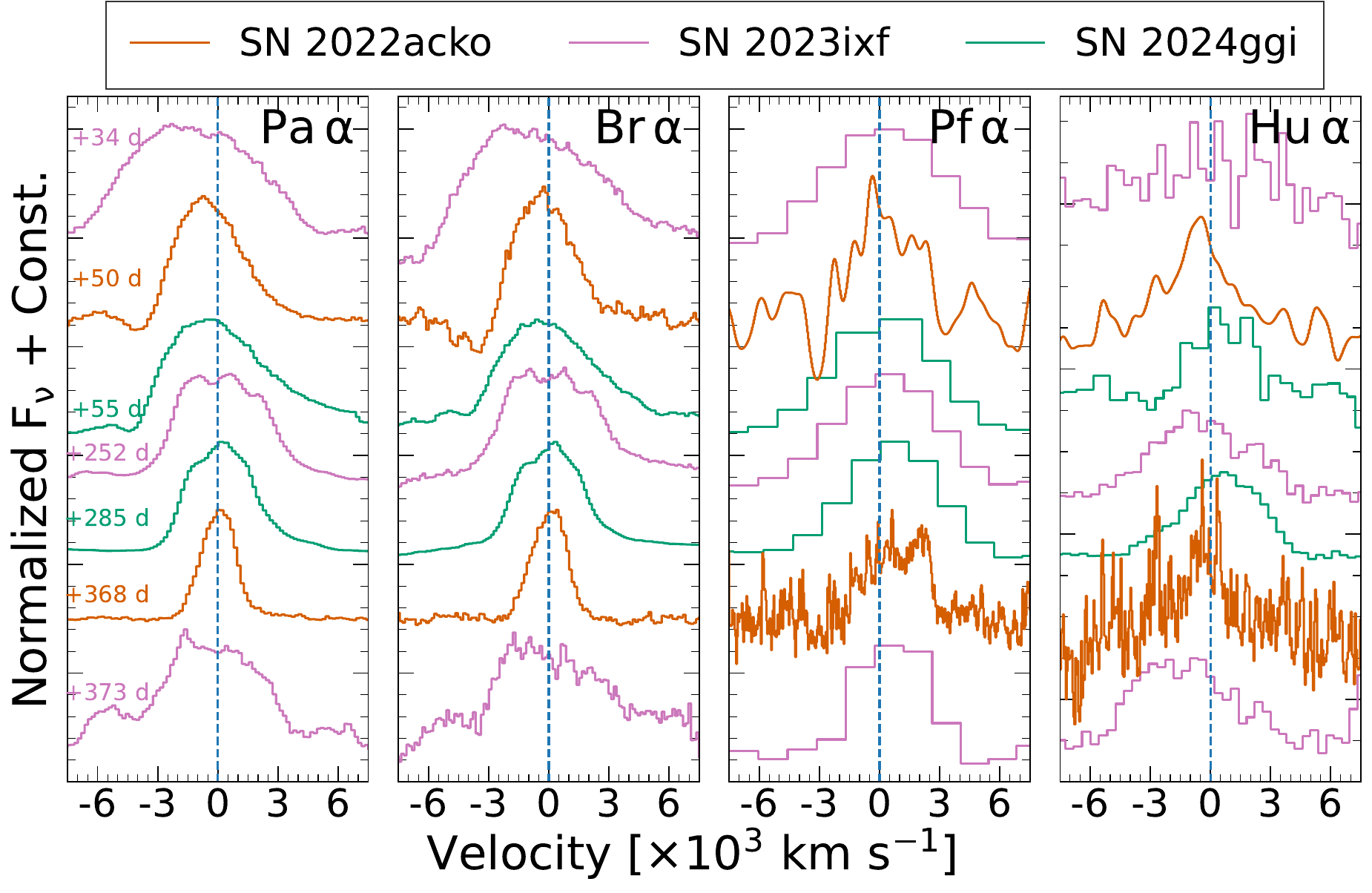}
    \caption{A comparison between the strong H features between \acko, SN\,2023ixf, and SN\,2024ggi.  \acko\ shows narrower emission line profiles than the other two SNe\,II. Additionally, \acko\ maintains a sharp peak in its H emission profiles, suggesting little to no dust was forming within the ejecta of \acko. }
    \label{fig:H_line_comp}
\end{figure*}

\begin{figure}
    \centering
    \includegraphics[width=\linewidth]{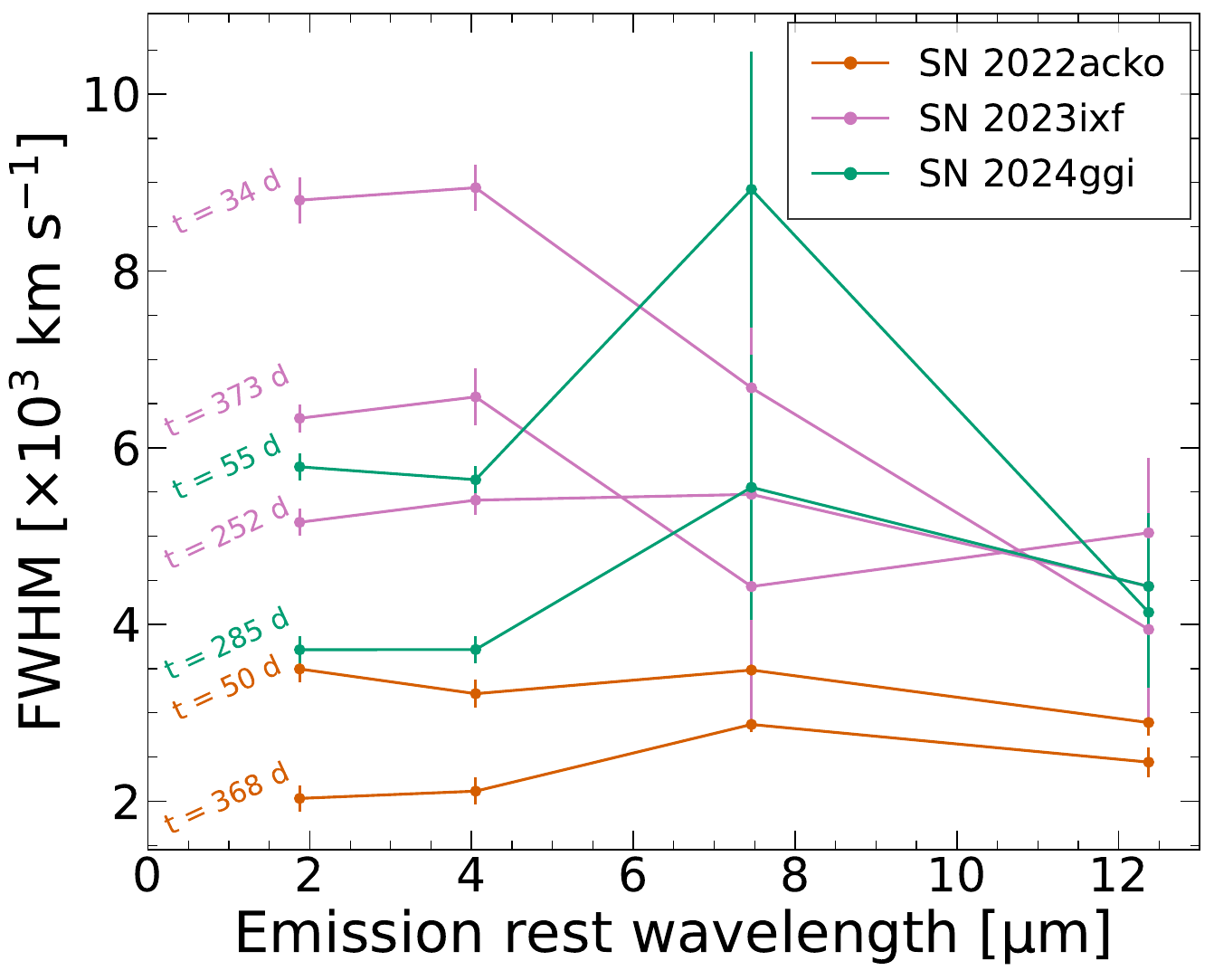}
    \caption{Evolution of the FWHM of the strong H alpha lines from the Paschen, Brackett, Pfund, and Humphreys series of \acko, SN\,2023ixf, and SN\,2024ggi. \acko\ is consistently slower than SN\,2023ixf and SN\,2024ggi. }
    \label{fig:H_FWHM_evo}
\end{figure}

\begin{deluxetable}{cccc}
    \tablecaption{FWHM and offset peak velocities for the strong emission lines shown in Figure~\ref{fig:line_vels}. Forbidden transitions are denoted by the square brackets around the emission wavelength. These velocities have not been corrected for the peculiar velocity at the location of \acko\ within NGC~1300, which has a value of $179.4$~\kms}
    \label{tab:FWHM_vs_Peak}
    \tablehead{\colhead{Emission Wavelength} & \colhead{Phase}& \colhead{FWHM}& \colhead{Offset velocity} \\
    \colhead{$[\mu m]$} & \colhead{[Days]} & \colhead{[\kms]} & \colhead{[\kms]}
    }
    \startdata
    \multicolumn{4}{c}{H} \\
    \hline
    $\mathrm{Pa}_\alpha$ & $367$ & $ 2000 \pm 200$ & $43 \pm 6$ \\
    $\mathrm{Br}_\alpha$ & $367$ & $2100 \pm 200$ & $33 \pm 5$ \\
    $\mathrm{Pf}_\alpha$ & $367$ & $2900 \pm 100$ & $491 \pm 31$ \\
    $\mathrm{Hu}_\alpha$ & $367$ & $2400 \pm 200$ & $-386 \pm 267$ \\
    \hline
    \multicolumn{4}{c}{He} \\
    \hline
    $1.083$ & $295$ & $1800 \pm 100$ & $140 \pm 11$ \\
    $1.083$ & $367$ & $1700 \pm 100$ & $159 \pm 35$ \\
    $2.058$ & $295$ & $1900 \pm 100$ & $454 \pm 58$ \\
    $2.083$ & $367$ & $1500 \pm 200$ & $277 \pm 90$ \\
    \hline
    \multicolumn{4}{c}{Neon} \\
    \hline
    $[12.813]$ & $295$ & $1000 \pm 400$ & $413 \pm 192$ \\
    $[12.813]$ & $367$ & $700 \pm 100$ & $490 \pm 40$ \\
    \hline
    \multicolumn{4}{c}{Argon} \\
    \hline
    $[6.985]$ & $295$ & $700 \pm 200$ & $718 \pm 79$ \\
    $[6.985]$ & $367$ & $900 \pm 200$ & $259 \pm 77$ \\
    \hline
    \multicolumn{4}{c}{Iron} \\
    \hline
    $1.980$ & $295$ & $2100 \pm 200$ & $233 \pm 76$ \\
    $1.980$ & $367$ & $1700 \pm 400$ & $1 \pm 173$ \\
    $[3.377]$ & $367$ & $2100 \pm 400$ & $201 \pm 163$ \\
    $[17.936]$ & $295$ & $900 \pm 200$ & $371 \pm 57$ \\
    $[17.936]$ & $367$ & $1100 \pm 400$ & $227 \pm 113$ \\
    \hline
    \multicolumn{4}{c}{Cobalt} \\
    \hline
    $[3.286]$ & $367$ & $1600 \pm 400$ & $466 \pm 165$ \\
    $[10.521]$ & $295$ & $1000 \pm 100$ & $413 \pm 40$ \\
    $[10.521]$ & $367$ & $1000 \pm 100$ & $286 \pm 27$ \\
    \hline
    \multicolumn{4}{c}{Nickel} \\
    \hline
    $[3.120]$ & $367$ & $1100 \pm 100$ & $105 \pm 21$ \\
    $[6.636]$ & $295$ & $1700 \pm 100$ & $218 \pm 28$ \\
    $[6.636]$ & $367$ & $1500 \pm 100$ & $235 \pm 11$ \\
    $[12.729]$ & $295$ & $1500 \pm 700$ & $363 \pm 211$ \\
    $[12.729]$ & $367$ & $1800 \pm 1400$ & $470 \pm 558$ \\
    \enddata
\end{deluxetable}

\section{Spectral velocities} \label{sec:vel}
Studying the properties of specific elements within the SED of \acko\ provides valuable insight into the distribution of material in the ejecta and the physics of the line-forming regions. This is effectively accomplished by analyzing the emission features of elements such as H, He, and Fe, which are strong features even in low-luminosity SNe\,II \citep{Reguitti_2021}.

\subsection{He, IME and IGEs}
We first examine the velocity structure of lines in the IMEs and IGEs identified in the $+295$ and $+368$~d spectra of \acko, which are presented in Figure~\ref{fig:line_vels}. 
The emission velocity profiles show that elements such as He, Fe and Ni are located near their rest wavelength with an average shift of $\sim +100$~\kms\ from $0$~\kms, suggesting a slightly aspherical distribution of material. Elements such as Ne and Ar display an offset from rest frame by $\sim +300$~\kms. As \acko\ evolved, the majority of the emission lines increased in strength relative to the continuum between the two epochs, reflecting ongoing changes in the ionization and density structure, with the most significant changes occurring in the [\ion{Ne}{2}]~$12.813$~\micron\ which appears to become more flat-topped and the [\ion{Ar}{2}]~$6.985$~\micron\ and [\ion{Fe}{2}]~$17.936$~\micron\ lines which become significantly stronger by day $+368$, emerging clearly in the later spectrum.
To understand the distribution of the ejecta within \acko\ we analyze the FWHM and offset from the rest frame wavelength of several emission lines from material located throughout the ejecta. To do this we fit the emission features shown in Figure~\ref{fig:line_vels} with a Gaussian function, masking regions that are contaminated by other emission features. For the elements with multiple strong emission lines, such as He, and Fe, we average the FWHM and offset peak for each epoch. In addition, we calculate the peculiar velocity of NGC~1300 at the radius of \acko, $\sim 4.6$~kpc from the galactic core, using eqn.~10 from \citet{Lang_2020} which gives a peculiar velocity of $179.4$~\kms. Assuming that the motion of \acko\ within its host galaxy was purely perpendicular to the line of sight, and thus moving at the maximum value of the peculiar velocity for its orbit, we subtract from the offset velocity derived from the Gaussian fitting. 
The average results of fitting the strong emission features of the IME and IGEs shown in Figure~\ref{fig:line_vels} at $+295$~d and $+368$~d are, after correcting for the peculiar velocity of \acko\ in NGC~1300, shown Figure~\ref{fig:Elem_FWHM_vs_Peak}. The results from this fitting for the individual emission lines are given in Table~\ref{tab:FWHM_vs_Peak}. A general trend can be observed where the material produced during the progenitor's evolution, i.e. the stellar burning products as Ar and Ne, show a narrower FWHM ($\sim 0.8 \pm 0.2 \times 10^3$~\kms) and larger offset velocity ($\sim 300 \pm 150$~\kms) than the emission features associated with the nuclear burning products synthesized by the explosion, e.g. Fe, Co, and Ni, which have a faster FWHM ($1.3 \pm 0.3 \times 10^3$~\kms) and smaller offset velocity ($\sim 100 \pm 100$~\kms). The only outlier from the outer elements is the He emission features which have a high average FWHM ($\sim 1.7 \pm 0.1 \times 10^{3}$~\kms) and lower offset velocity ($\sim 75 \pm 25$~\kms). The differences between the He, IME and IGEs suggest a non-spherical distribution of material within \acko\ as it exploded. We discuss the repercussions of this ejecta structure in Section~\ref{sec:structure}.

\subsection{Hydrogen}
The most prominent emission lines identified within the \jwst\ spectra of \acko\ are those associated with different H transitions. For most of the H series, the $\alpha$ through $\delta$ transitions are clearly identified in the \jwst\ spectra of \acko, with additional higher order transitions observed in the MIRI/MRS data, e.g. Hu$_{15 \rightarrow 6}$. The strongest transitions in the spectra are the $\alpha$-transitions, which can be used to study the velocity distribution of the H envelope in \acko. The average FWHM velocities of the H series $\alpha$-transitions in the $+369$~d \jwst\ spectrum of \acko\ are given in Table~\ref{tab:FWHM_vs_Peak}. The Gaussian fitting clearly shows that the bulk of the H in \acko\ is contained within $2,400 \pm 100$~\kms, showing a significant reduction in velocity from the initial \jwst\ observation \citep{Shahbandeh_2024}. Additionally, the H lines tend to be located around $0$~\kms, with the exception of Pf$_\alpha$ and Hu$_\alpha$, which are heavily blended with other emission lines. The low offset velocity of the H emission lines suggest that the H envelope of \acko\ was quite spherical.

We look at the $\alpha$ transitions of the H Paschen, Brackett, Pfund, and Humphreys series in \acko\ and compare them with the same lines identified in SN\,2023ixf \citep{Medler_2025, DerKacy_2026} and SN\,2024ggi \citep{Dessart2025, Baron_2025, Mera_2026}. The profiles of the three SNe at comparable epochs are shown in Figure~\ref{fig:H_line_comp}. 
All emission features have been continuum subtracted by randomly sampling the spectral continuum surrounding the emission profile to obtain an average continuum which was then subtracted to remove contamination from blended features and dust emission within the spectrum. After continuum subtracting the emission features, we fit a Gaussian function to each emission profile in velocity space to determine the FWHM of each profile. For SN\,2023ixf, we fit the central core of the H emission features and ignore the high-velocity wings identified in the earlier spectra \citep{Li_2025, Medler_2025} to focus on the velocity evolution of the bulk H-rich material. The FWHM results of these fits are shown in Figure~\ref{fig:H_FWHM_evo} as a function of the emission lines rest wavelength.

\begin{figure*}[t]
    \centering
    \includegraphics[width=0.95\linewidth]{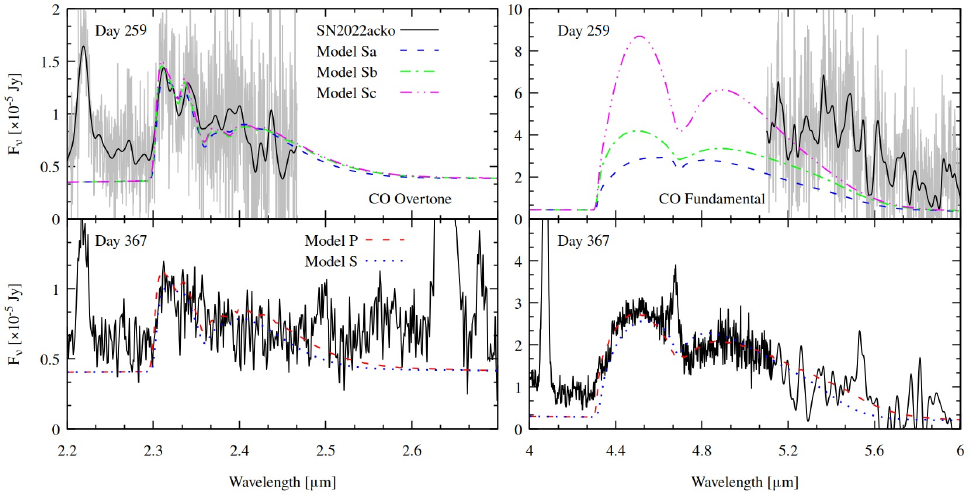}
    \caption{Comparing our prolate (P) and non-clumping (S) models to the observations of \acko. The upper panels show our results at 259 days, and the lower panels at 367 days. Without the modes of the fundamental band, we restrict our fits at day 259 to non-clumping models and show the sensitivity of the temperature profile (see Figure \ref{fig:temp}). }
    \label{fig:molecule_fit}
\end{figure*}

\acko\ displays distinct emission features that distinguish it from SN\,2023ixf and SN\,2024ggi. Even during the photospheric phase at $+50$~d, \acko\ possesses lower spectral velocities than SN\,2023ixf by $\sim 5500$~\kms\ at the bluer wavelengths and $\sim 1000$~\kms\ at redder wavelengths \citep{DerKacy_2026}. The H emission features in the photospheric phase spectrum of \acko\ are quite similar in shape to those of SN\,2024ggi \citep{Baron_2025} at $+55$~d peaking around their rest wavelengths. Despite the similarities in shape the H features of \acko\ are $\sim 2000$~\kms\ narrower than SN\,2024ggi, see Figure~\ref{fig:H_FWHM_evo}, a result of \acko's kinetic energy being smaller \citep{Hamuy_2002}. While more similar to SN\,2024ggi than SN\,2023ixf, the H features in \acko\ lack the faster wings seen in SN\,2024ggi. We observe a decrease in the FWHM of the H profiles at longer wavelengths in all SNe, although this trend is stronger in SN\,2023ixf and SN\,2024ggi than in \acko. The stronger trend observed in SN\,2023ixf and SN\,2024ggi is likely due to the lower resolving power of the their \jwst\ observations \citep{Baron_2025, DerKacy_2026} compared to those of \acko, as well as the blending of other non-resolved lines, such as Pf$_\alpha$ blending with the Hu$_\beta$ emission line. Later, during the nebular phase \acko\ displays a narrow sharp peaked emission feature whereas both SN\,2023ixf and SN\,2024ggi show complex structure in the peak of their emission features, Figure~\ref{fig:H_line_comp}. In SN\,2023ixf, the broad flat-topped shape can be accounted for by the presence of dust within the system attenuating the redward flux of the profile \citep{Bevan_2016, Medler_2025}. Whereas in SN\,2024ggi, the observed blue notch and change in the slope on the red side of the line profiles may indicate the start of dust formation \citep{Bevan_2016}. This strongly suggests that \acko\ does not form dust on the timescales observed in other H-rich CC-SNe, which may be explained by the smaller ejecta mass and kinetic energy of \acko\ compared to typical SNe\,II resulting in a smaller dust formation region within the ejecta. 

\section{CO formation in the SN ejecta} \label{sec:CO_form}
To model the molecular features of \acko\ we employed the MOlecular Fitting Analysis Tool \citep[MOFAT:][]{2026mera_theory}, perviously used to model the CO emission in SN\,2024ggi \citep{Mera_2026}. MOFAT is a seven-parameter iterative framework that fits observed molecular-band fluxes to identify optimal parameter set solutions. MOFAT determines the total amount of CO mass (${M_{CO}}$) required to reproduce the CO emission features by taking into account the temperature (\textit{T$_1$}) at the inner edge of the CO shell defined at a specific velocity (\textit{v$_1$}), which has a velocity width of \textit{$\Delta$v}, and \textit{n} is the slope of the CO density distribution. Additionally, when clumps within the ejecta are taken into account MOFAT allows the relative size of the clumps with respect to the radius \textit{$r_R$}, the density enhancement factor within the clump with respect to its environment \textit{f$_c$}, and the flattening parameter of the spheroid shape of the clump, $\epsilon$, to be set. 
It assumes a spherically symmetric large-scale geometry with clumpy (spheroid) substructures and solves the corresponding radiative transfer equations. Molecular opacities are computed using HYDRA \citep{Hoeflich_1988, Sharp_1990, gerardy02, Rho18}, where MOFAT applies non-LTE corrections. However, rather than solving time-dependent molecular formation, MOFAT uses observations to constrain key parameters and infer the temperature structure and physical conditions at a given epoch.  

\begin{figure}
    \centering
    \includegraphics[width=0.95\linewidth]{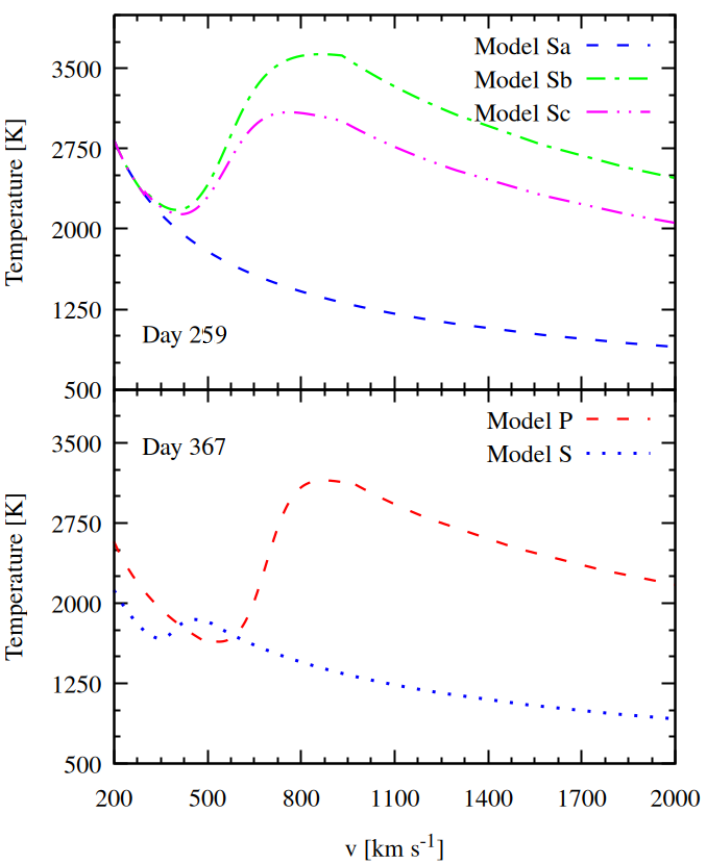}
    \caption{Temperature structures of optimized parameter sets with and without clumps. In the upper panel, we show day 259, where different temperature structures are shown to demonstrate their effect on the CO profiles. The bottom panel shows day 367. See Table \ref{tab:model_values} for the model parameters and Figure \ref{fig:molecule_fit} for their fits.}
    \label{fig:temp}
\end{figure}

\begin{deluxetable}{ccccccccccc}
    \caption{Parameter values from Figure \ref{fig:molecule_fit} optimized for the two epochs observed. Model P indicates the best-fitting models for prolate clumps, while Model S is without clumps.  The \textit{Clumps} column indicates whether the model takes into account clumping effects. The \textit{Mass} column shows the total amount of CO mass. \textit{T$_1$} shows the temperature at the inner edge (\textit{v$_1$}) of the CO shell, which has a width of \textit{$\Delta$v}. \textit{n} is the slope of the CO density distribution. \textit{$r_R$} is the relative size of the clumps with respect to the radius. \textit{f$_c$} is the density enhancement factor within the clump with respect to its environment. $\epsilon$ is the flattening parameter of the spheroid shape of the clump (when $\epsilon =1$ it's a sphere, and $>1$ a prolate).}
    \label{tab:model_values}
    \tablehead{\colhead{Model} & \colhead{Time [d$]$} & \colhead{Clumps} & \colhead{Mass [M$_\odot$$]$}& \colhead{T$_1$ [K$]$} & \colhead{v$_1$ [km/s$]$} & \colhead{$\Delta$v [km/s]} & \colhead{n} & \colhead{$r_R$} & \colhead{f$_c$} & \colhead{$\epsilon$}}
    \startdata
    P  & $368$ & yes & $6.81e-4$ & $1656$ & $505$ & $708$ & $9.33$ & $1.73e-1$ & $3.82$ & $11$ \\ 
    \hline 
    \multirow{2}{*}{S$_{\rm a,b,c}$} & $259$ & no & $1.55e-4$ & $2346$ & 300 & $1000$ & $5.00$ & - & - & - \\ 
    & $368$ & no & $2.47e-4$ & $1743$ & $300$ & $950$ & $10.00$ & - & - & - 
    \enddata
\end{deluxetable}

In Figure~\ref{fig:molecule_fit}, we show our best-fit solutions for the overtone band at $+259$ days and for both the overtone and fundamental bands at $+368$ days. At both epochs, we present non-clumped models (S), while for day $+368$ we additionally include a prolate clumping model, as the optically thick fundamental band is required to constrain clump parameters. The corresponding temperature profiles are shown in Figure~\ref{fig:temp}, with all model parameters listed in Table~\ref{tab:model_values}. 

At day 367, the clumpy model for the fundamental band provides an improved fit to the observations at the $\sim 10\%$ level, and this epoch offers the strongest constraint on the substructure of the ejecta. We find that the relative clump size ($r_R = 1.73\times10^{-1}$) is comparable to that inferred for SN~2024ggi at day 385 ($r_R \approx 1.5\times10^{-1}$) \citep{Mera_2026}, suggesting a similar characteristic scale. Such clump sizes may plausibly arise from Rayleigh--Taylor instabilities or neutrino-driven instabilities, although further comparison with multi-dimensional simulations \citep[e.g.][]{2009couch, 2019burrows, 2025vartanyan} is needed to assess this connection. The inferred clump morphologies are also similar, with $\epsilon = 11$ for \acko\ and $\epsilon = 14$ for SN~2024ggi. However, in the absence of constraints from earlier epochs for \acko, we cannot determine whether the clump sizes and shapes evolve with time. If we instead adopt the masses inferred from non-clumped (spherically symmetric) models, we find that the total CO mass increases with time ($1.55\times10^{-4}$ to $2.47\times10^{-4}$~\msolar), indicating that the ejecta may still be in an active phase of CO formation, consistent with trends observed in other SNe. We also note that the clumpy models yield systematically different CO masses compared to the spherically symmetric case, as the clumps approach a semi-optically thick regime in the overtone band ($f_c = 3.82$), which modifies the emergent flux and inferred mass.

Although clumpy models are required to accurately reproduce the molecular line profiles at later epochs, the spherically symmetric models still provide useful physical insight into the ejecta structure. From day 259 to 367, the molecular-forming region appears fully exposed, with the inner velocity remaining approximately constant at $v_1 \sim 300~\mathrm{km~s^{-1}}$ and a characteristic width of $\Delta v \sim 1000~\mathrm{km~s^{-1}}$. The growth of the CO mass is concentrated in the coolest regions of the ejecta, as indicated by an increase in the density profile index from $n \approx 5$ to $n \approx 10$. 
Lower-energy, lower-mass Type~II supernovae are expected to produce more centrally confined and compact carbon-oxygen layers (e.g., \citealt{woosley93, woosley95, woosley2002evolution}). Consistent with this picture, \cite{VanDyk_2023} infer a zero-age main sequence (ZAMS) mass of $\sim 8$ \msolar\ for \acko, whereas SN~2024ggi is estimated to have a higher ZAMS mass of $\approx 12$--$15$ \msolar\ \citep{2025ertini, 2025ferrari, 2026hueichapan}. Correspondingly, the molecular-forming region in SN~2024ggi is located at significantly higher velocities, with $v_1 \approx 1000~\mathrm{km~s^{-1}}$ and $\Delta v \approx 3000~\mathrm{km~s^{-1}}$. This comparison suggests that the extent of the CO-forming region may serve as a useful proxy for the progenitor ZAMS mass.

At day 259, we illustrate the effect of the temperature structure on the absolute flux levels of the fundamental band and demonstrate the sensitivity of the spectrum to variations in the temperature profile. In \citet{2026mera_theory}, the temperature profile is constrained by boundary conditions set by the absolute fluxes of both the overtone and fundamental bands. We consider three representative temperature structures for Model S. The three models are shown in Figure~\ref{fig:temp} and are labeled as Models Sa, Sb, and Sc. Model Sa corresponds to a temperature profile without reheating of the outer molecular ejecta; in this case, the optical depth of the fundamental band is extremely large, which suppresses its volumetric emission. As a result, the flux from both modes originates from approximately the same radius, producing a relatively flat-topped profile. In contrast to Model Sa, the second model, Model Sb, assumes significant reheating of the outer ejecta by the photon field, which reduces the opacity of the fundamental band in these regions (see Figure~18 in \citealt{Rho21}). This reduction lowers the absolute flux contribution from the outer layers, since the emergent intensity scales as $1 - \exp(-\tau)$ and thus drops with decreasing $\tau$. Consequently, an intermediate temperature structure between Model Sa and Sb is required to reconcile these effects. This is achieved in Model Sc, which provides a balance between the two extremes and successfully reproduces the observed absolute flux levels near $5.2\,\micron$. 

The temperature profiles at day 367 differ significantly between the non-clumped and clumped models due to the interplay between molecular formation, cooling, and radiative transfer. In the non-clumped case, the continued growth of molecules implies that CO is actively forming and cooling the gas, establishing a feedback cycle in which enhanced molecular formation leads to increased cooling, which in turn promotes further molecule formation. This behavior is most evident near the velocity minimum at $\approx 300~\mathrm{km~s^{-1}}$, where the accumulation of molecules increases the local mass and steepens the density profile (i.e., increases $n$). As a result, the outer regions become progressively more optically thin and begin to follow the radiation field, asymptotically approaching a $T \propto r^{-1/2}$ profile. In contrast, the clumped models exhibit a markedly different behavior due to the presence of optically thick clumps ($f_c = 3.8$). In this regime, the clumps in the outer regions absorb a larger fraction of the radiation field, leading to a partial reheating of the outer envelope, resulting in a temperature structure that deviates from the non-clumped case.



\begin{figure}
    \centering
    \includegraphics[width=\linewidth,  trim={3cm, 2cm, 3cm, 0cm}, clip=True]{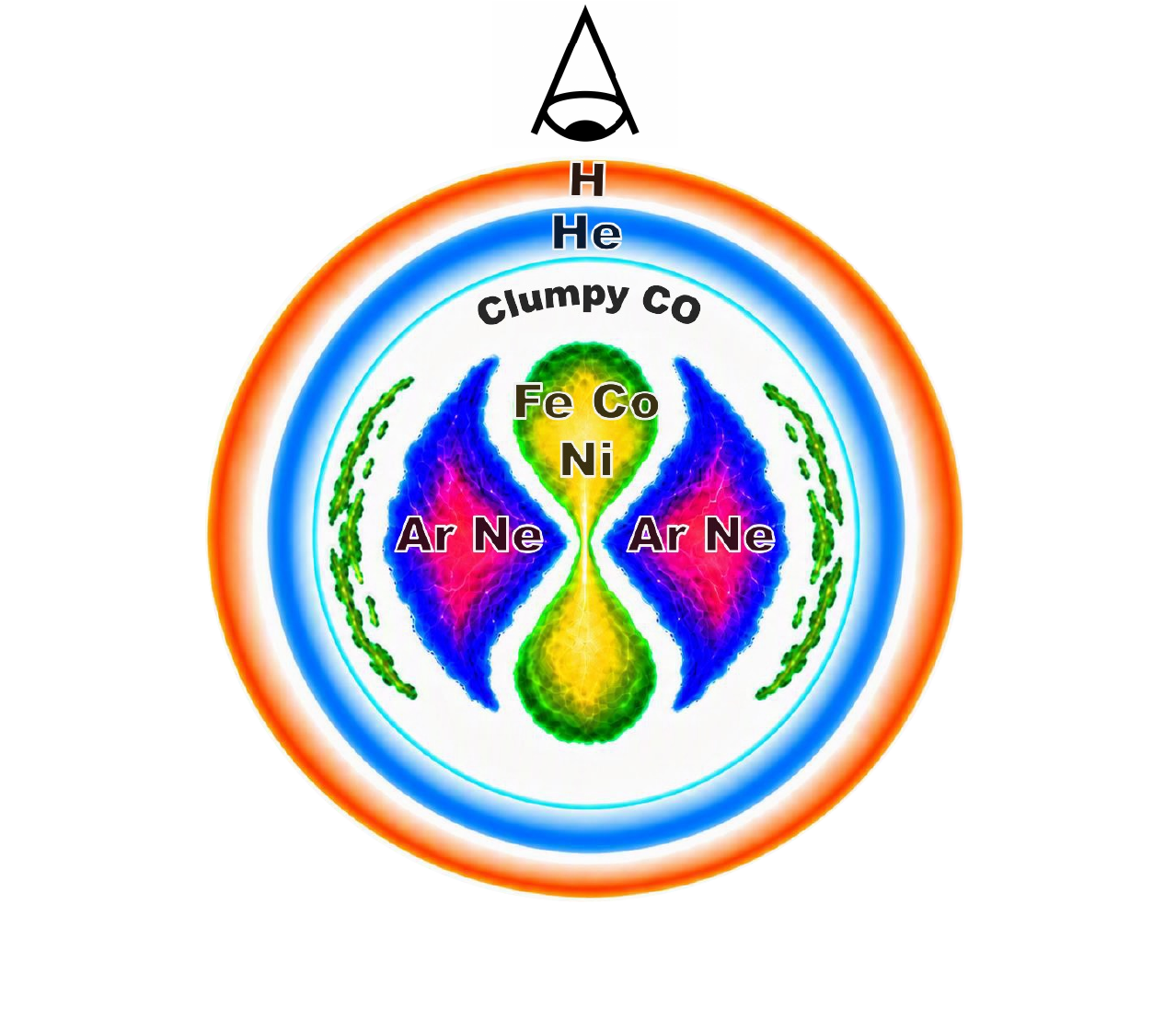}
    \caption{Structure of \acko\ just after the explosion required to explain the velocity distribution of the He, IME and IGE shown in Figure~\ref{fig:Elem_FWHM_vs_Peak}. Note that the H and He shell is far enough from the central explosion to not be affected by the asphercities in the inner ejecta.  }
    \label{fig:acko_structure}
\end{figure}

\section{Ejecta Distribution and Instability}\label{sec:structure}

The emission-line profiles of H, He, IMEs, and IGEs reveal that \acko\ cannot be fully described by a simple spherically symmetric SN\,II explosion.
Given the difference between the velocity offsets of the stellar burning products and the nuclear burning products, we suggest an asymmetric explosion. The origin of this asymmetry is likely from a bipolar wind that is launched from an accretion disk formed around the compact central object from in-falling material \citep{Hoeflich_2001, Hoeflich_2004}. This bipolar wind expands outwards along the axis of rotation of the compact object, displacing stellar burning products along its path \citep{Hoeflich_2001}, resulting in a structure where the IGEs are predominantly along the line of sight and the IMEs are located perpendicular in the ejecta. As a bipolar wind is predicted to fail prior to it breaking through the stellar surface \citep{Hoeflich_2004}, the outer H and He envelopes are left largely unaffected by the motion of the inner layers of \acko's progenitor. This origin for the asymmetry agree with the velocity distribution of H and He observed in the NIR and MIR observations of \acko. 

Further asymmetry could be produced by Rayleigh-Taylor instabilities, which are expected to extend out to the carbon-oxygen layer and He envelope interface \citep{Kifonidis_2000, Kifonidis_2003}. In such a scenario, the bulk of the nuclear burning products are either expelled along the line of sight with plumes of IGE moving into the outer regions throughout the star resulting in Rayleigh-Taylor instabilities at the interface between different density gradients. The enhancement of the asymmetry brought about by the Rayleigh-Taylor instabilities can explain the clumping we require for modeling the CO emission features observed in the \jwst\ spectra of \acko. A similar distribution of material was observed in SN\,1987A \citep{Wang_2001}, as well as in several other SNe via polarimetric observations \citep{Wang_1996, Maund_2010}. 

An alternative method to create the observed asymmetry in IME and IGEs is a choked jet that fails to pierce the He envelope resulting in the mixing of the IME regions \citep{Khokhlov_1999, Couch_2009, Suzuki_2022}, however the choked jet mechanism is not favored as current models require a neutrino driven explosions \citep{Burrows_2024}. 

Based on the velocity observations and mechanisms discussed above, we suggest that the structure of \acko\ shown in Figure~\ref{fig:acko_structure} may best explain the distribution of the IGE-rich inner ejecta, the IME-rich outer ejecta and the H/He outer envelope we observe from the NIR and MIR emission lines. We do not constrain the properties of the S/Si layer as there are not enough isolated lines to determine the FWHM and offset velocity for elements within this layer.   

Alongside the different velocity distributions of the IMEs and IGEs, which can be explained by chemical asymmetries produced by a bipolar wind or outflow, we also observe a systematic offset of the IME and IGE emission-line peaks from zero velocity, as shown in Figure~\ref{fig:Elem_FWHM_vs_Peak}. The median velocity offset for the elements discussed in Section~\ref{sec:vel} is $97.4^{+86.3}_{-42.3}$~\kms, where the uncertainties are determined by randomly sampling the measured range of offset velocities $1000$ times and taking the 16th and 84th percentiles. Importantly, this offset does not necessarily require the individual line profiles to be strongly asymmetric; the profiles may remain approximately symmetric around their own shifted centroids. Instead, the systematic shift indicates that the line-emitting material is displaced in velocity space by $\sim100$~\kms\ relative to the adopted rest frame. Because we account for the peculiar velocity of \acko\ within the host galaxy, this displacement cannot be explained by residual uncertainty in the local systemic velocity. If the displacement reflects a true net asymmetry in the ejecta momentum, then momentum conservation implies that the compact remnant should receive a natal kick in the opposite direction to the net ejecta motion, as expected for a newly formed neutron star \citep{Janka_1994, Burrows_1996, Scheck_2006}. Thus, the systematic offset in the emission-line peaks observed in \acko\ supports the presence of a neutron-star remnant.

\section{Conclusions} \label{sec:conc}
\acko\ was the first SN\,II to be spectroscopically observed with \jwst, and remains the only SN\,II observed with \jwst\ MIRI/MRS during both the plateau and nebular phases. The initial analysis of this event placed an upper limit of $<10^{-8}$~\msolar\ on the CO mass and found no evidence for SiO emission \citep{Shahbandeh_2024}. In this work, we extend the study of \acko\ to later phases using \jwst\ spectra obtained at $+259$ and $+368$~d after explosion. These observations provide a unique view of the transition from the plateau to the nebular phase in a low-luminosity SN~II, allowing us to follow the emergence of IMEs, IGEs, and molecular emission in the infrared.

The spectra of \acko\ show clear evolution from the initial \jwst\ epoch. At later phases, the infrared spectra are dominated by strong H emission together with forbidden lines from ionized Ar, Ne, Fe, Co, and Ni. The emission profiles show that most of the emitting material is confined to velocities below $\sim 2000$~\kms. The exception is H, which has a characteristic velocity of $2400 \pm 100$~\kms, although this is still $\sim 1000$--$2000$~\kms\ slower than in the initial \jwst\ observation. Compared to SN~2023ixf and SN~2024ggi, \acko\ is substantially narrower in H, with a FWHM approximately one third that of SN~2023ixf and one half that of SN~2024ggi. This difference is consistent with \acko\ being a lower-energy explosion from a lower-mass progenitor.

We modeled both the CO overtone and fundamental emission in \acko\ at $+259$ and $+368$~d using the Molecular Fitting and Analysis Tool, MOFAT. Spherically symmetric models reproduce the global properties of the molecular emission, but clumpy models are required at later times to capture the detailed line profiles. At $+368$~d, the inclusion of clumping improves the fit by $\sim 10\%$ and implies a characteristic clump size of $r_R \sim 0.17$ and a highly prolate morphology with $\epsilon \sim 11$. The inferred CO mass increases from $1.55\times10^{-4}$~\msolar\ at $+259$~d to $2.47\times10^{-4}$~\msolar\ at $+368$~d, demonstrating ongoing molecule formation in the ejecta. The CO-emitting region is centrally concentrated, with $v_1 \sim 300$~\kms\ and $\Delta v \sim 1000$~\kms, consistent with expectations for a low-mass progenitor of $\sim 8$~\msolar. These results show that CO emission is a sensitive probe of the thermal evolution, geometry, and small-scale structure of SN ejecta, and that the extent of the molecule-forming region may provide a diagnostic of progenitor mass.

Despite the clear emergence of CO emission, \acko\ shows little to no continuum emission associated with warm dust. This makes \acko\ distinct from SN~2023ixf, where warm dust emission was detected at comparable phases \citep{Medler_2025}. The lack of pre-existing dust is consistent with HST constraints on the progenitor environment \citep{VanDyk_2023}, while the weak dust continuum and lack of SiO emission suggest that freshly formed dust is also limited at these epochs. Thus, \acko\ shows that low-luminosity SNe~II can form molecules, but may not efficiently convert these molecules into dust on the same timescales as more massive SNe~II. Any associated dust formation may therefore occur only on longer timescales, if at all.

The line profiles also reveal evidence for large-scale asymmetry in the ejecta of \acko. The IGE features, together with the H and He envelopes, peak close to rest wavelength but share a systematic offset of $\sim +100$~\kms, while IMEs such as Ar and Ne are offset by $\sim +300$~\kms. This difference suggests that the IME-rich layers are distributed differently from both the inner IGE-rich material and the outer H/He envelope, pointing to an asymmetric chemical structure in the ejecta of \acko. We interpret this chemical asymmetry as evidence that the stellar-burning products were displaced or mixed during the explosion, potentially by fallback accretion onto the newly formed compact remnant and the associated bipolar wind or outflow. Rayleigh-Taylor instabilities at the boundaries between compositional layers may have further enhanced this mixing and asymmetry in the IME-rich ejecta.

In addition to this chemical asymmetry, the emission-line peaks show a common residual velocity offset of $\sim100$~\kms\ after accounting for the local host-galaxy velocity field. This bulk offset is distinct from the larger IME displacement and provides evidence for the ejecta-side momentum signature of the natal kick imparted to the newly formed neutron star. In this picture, asymmetric mass ejection gives the compact remnant a recoil velocity in the opposite direction through momentum conservation. The magnitude of the offset is comparable to the natal velocities of slow-moving NSs in the Milky Way, supporting the presence of a NS remnant in \acko. Thus, the chemically distinct IME distribution, the fallback-powered bipolar outflow, and the bulk velocity offset are linked observational signatures of the same asymmetric explosion.

Taken together, these observations show that \acko\ provides a rare view of the chemical and structural evolution of a low-mass SN~II. The high spectral resolution and broad infrared coverage of \jwst\ MIRI/MRS make it possible to isolate IME and IGE lines that are inaccessible from the optical alone, while the detection of CO emission reveals the onset of molecule formation in the inner ejecta. At the same time, the weak dust continuum and lack of SiO emission suggest that low-mass SNe~II may contribute less efficiently to early dust production, or may form dust on longer timescales than more massive events. Future late-time \jwst\ observations of low-luminosity SNe~II will be essential for determining how common this behavior is, how their ejecta are structured, and how these numerous explosions contribute to the chemical enrichment and dust budgets of galaxies.

\vspace{0.5cm}
\begin{acknowledgements}
K.M., T.M, E.B., C.A., J.D., M.S., and P.H., acknowledge support from NASA grants JWST-GO-02114,
JWST-GO-02122, JWST-GO-04522, JWST-GO-04217, JWST-GO-04436,
JWST-GO-03726, JWST-GO-05057, JWST-GO-05290, JWST-GO-06023,
JWST-GO-06677, JWST-GO-06213, JWST-GO-06583. Support for
programs \#2114, \#2122, \#3726, \#4217, \#4436, \#4522,  \#5057,
\#6023, \#6213, \#6583, and \#6677
were provided by NASA through a grant from the Space Telescope Science
Institute, which is operated by the Association of Universities for Research in
Astronomy, Inc., under NASA contract NAS 5-03127.
Support for program \#2122 was provided  by NASA through a grant from the Space Telescope Science  Institute, which is operated by the Association of Universities for Research in Astronomy, Inc., under NASA contract NAS 5-03127.

P.H. acknowledges support by the National Science Foundation (NSF) through grant AST-2306395 and NASA grant 80NSSC20K0538.
E.B. acknowledges support by NASA grant 80NSSC20K0538. 
M.D.S. is funded by the Independent Research Fund Denmark (IRFD, grant number  10.46540/2032-00022B). 
L.G. acknowledges financial support from CSIC, MCIN and AEI 10.13039/501100011033 under projects PID2023-151307NB-I00, PIE 20215AT016, CEX2020-001058-M, and by the MaX-CSIC Excellence Award MaX4-SOMMA-ICE.
Y.Y.'s research is partially supported by the Tsinghua University Dushi Program.
J.L. is supported by grant NSF-2206523.
M.A.T. acknowledges support from Program number HST-GO-17429.001-A with funding provided through a grant from the STScI under NASA contract NAS5-26555.
A.D. is supported by the European Research Council (ERC) under the European Union’s Horizon 2020 research and innovation programme under Grant Agreement No. 101002652 (BayeSN; PI K. Mandel) and Marie Skłodowska-Curie Grant Agreement No. 873089 (ASTROSTAT-II).

This work is based on observations made with the NASA/ESA/CSA James Webb Space Telescope. 
The data were obtained from the Mikulski Archive for Space Telescopes at the Space Telescope Science Institute, which is operated by the Association of Universities for Research in Astronomy, Inc., under NASA contract NAS 5-03127 for JWST. These observations are associated with program \#2122. The specific observations analyzed in this work can be accessed via 
\end{acknowledgements}

\facilities{JWST (MIRI/MRS,NIRSpec), MAST (JWST), Keck-II (NIRES), GTC (EMIR)}

\software{jwst (\citealp[ver. 1.18.0,][]{Bushouse_2025_JWSTpipeline}), jdaviz (\citealp[ver. 3.2.1,][]{jdaviz}), Spextool \citep{Cushing_2004_Spextool}, HYDRA \citep{Hoeflich2003,Hoeflich2009,Hoeflich_etal_2017},
MOLFAST \citep{2026mera_theory}, OpenDx (an open-sourced visualization package developed by IBM), Astropy \citep{astropy:2013, astropy:2018, astropy:2022}, NumPy \citep{numpy2020}, SciPy \citep{SciPy2020}, Matplotlib \citep{matplotlib}, pypeit \citep{2020JOSS....5.2308P}. }

\clearpage
\bibliographystyle{aasjournal}
\bibliography{refs}

\appendix


\begin{deluxetable}{lccc}
  \renewcommand\thetable{A.1}
  \tablecaption{\acko\ spectral log for the second epoch of observations
  \label{tab:info_ep2}} 
  \tablehead{\colhead{Parameter} & \colhead{Value}& \colhead{Value}& \colhead{Value}}
  \startdata
    \multicolumn{4}{c}{MIRI/MRS Spectra}  \\
        \hline
    Grating& Short &Medium & Long\\
    Groups per Integration& 36&36&36 \\
    Integrations per Exp. &1&1&1\\
    Exposures per Dither &1&1&1\\
    Total Dithers & 4 & 4& 4\\
    Exp Time [s$]$ & 10511.564 & 10511.564 & 10511.564 \\
    Resolution\tablenotemark{a} &  \multicolumn{3}{c}{$\sim$2,700}\\ 
    Epoch\tablenotemark{a} [days$]$ &  \multicolumn{3}{c}{258.79}\\ 
    $T_\text{obs}$ [MJD$]$ & \multicolumn{3}{c}{60178.32} \\  
    \hline
    \enddata
  \tablecomments{~\tablenotemark{a}~Rest frame days relative to 
  time of explosion, MJD $=59918.17$ \citep{Bostroem23}.}
\end{deluxetable}

\begin{deluxetable}{lccc}
  \renewcommand\thetable{A.2}
  \tablecaption{\acko\ spectral log of the third \textit{JWST} observational epoch 
  \label{tab:info_ep3}} 
  \tablehead{\colhead{Parameter} & \colhead{Value}& \colhead{Value}& \colhead{Value}}
  \startdata
    \multicolumn{4}{c}{MIRI/MRS Spectra}  \\
        \hline
    Grating& Short &Medium & Long\\
    Groups per Integration& 36&36&36 \\
    Integrations per Exp. &1&1&1\\
    Exposures per Dither &1&1&1\\
    Total Dithers & 4 & 4& 4\\
    Exp Time [s$]$ & 7166.98 & 7166.98 & 7166.98 \\
    Resolution\tablenotemark{a} &  \multicolumn{3}{c}{$\sim$2,700}\\ 
    Epoch\tablenotemark{a} [days$]$ &  \multicolumn{3}{c}{367.67}\\ 
    $T_\text{obs}$ [MJD$]$ & \multicolumn{3}{c}{60287.77} \\  
    \hline
    \multicolumn{4}{c}{NIRSpec Acquisition Image}  \\
    \hline
    Filter & \multicolumn{3}{c}{CLEAR} \\
    Exp Time [s$]$ & \multicolumn{3}{c}{0.08} \\
    Readout Pattern & \multicolumn{3}{c}{NRSRAPID} \\
    \hline
    \multicolumn{4}{c}{NIRSpec Spectra}  \\
    \hline
    Slit & \multicolumn{2}{c}{S400A1} & S400A1 \\
    Filter & \multicolumn{2}{c}{F170LP} & F290LP \\
    Grating & \multicolumn{2}{c}{G235M} & G395M \\
    Groups per Integration & \multicolumn{2}{c}{10} & 10 \\
    Integrations per Exp. & \multicolumn{2}{c}{2} & 2 \\
    Exposures per Dither & \multicolumn{2}{c}{1} & 1 \\
    Resolution &\multicolumn{3}{c}{1000}\\
    Total Dithers & \multicolumn{2}{c}{3} & 3 \\
    Exp Time [s$]$ & \multicolumn{2}{c}{383.39} & 383.39 \\
    Epoch\tablenotemark{a} [days$]$ & \multicolumn{2}{c}{367.67}  & 367.67 \\ 
    $T_\text{obs}$ [MJD$]$ & \multicolumn{2}{c}{60287.77} & 60287.77 \\
      \hline
  \enddata
  \tablecomments{~\tablenotemark{a}~Rest frame days relative to 
  time of explosion, MJD $=59918.17$ \citep{Bostroem23}.}
\end{deluxetable}

\begin{deluxetable}{cccc}
    \renewcommand\thetable{A.3}
  \tablecaption{\acko\ spectral log of the ground-based observations
  \label{tab:GB_info}} 
  \tablehead{\colhead{Telescope /} & \colhead{Observation date}& \colhead{Phase}& \colhead{Exp Time} \\
  \colhead{Instrument} & \colhead{[MJD]}& \colhead{[days]}& \colhead{[seconds]}}
  \startdata
  GTC/EMIR & 59951.99 & 33.6 & 1439.95 \\
  Keck/NIRES & 60212.53 & 292.8 & 9600 \\
  Keck/NIRES & 60314.30 & 394.1 & 9600
  \enddata
\end{deluxetable}

\end{document}

%% file: authors.tex
\author[0000-0002-5221-7557]{K.~Medler}
  \email{kmedler@hawaii.edu}
 \affiliation{Institute for Astronomy, University of Hawaiʻi at M\=anoa, 2680 Woodlawn Dr., Hawaiʻi, HI 96822, USA }
 
\author[0000-0001-5888-2542]{T.~Mera}
  \email{tycomera@gmail.com}
   \affiliation{Department of Physics, Florida State University, 77 Chieftan Way, Tallahassee, FL 32306, USA}
 \affiliation{Institute for Astronomy, University of Hawaiʻi at M\=anoa, 2680 Woodlawn Dr., Hawaiʻi, HI 96822, USA }

 \author[0000-0002-5221-7557]{C.~Ashall}
  \email{cashall@hawaii.edu}
 \affiliation{Institute for Astronomy, University of Hawaiʻi at M\=anoa, 2680 Woodlawn Dr., Hawaiʻi, HI 96822, USA }

 \author[0000-0002-4338-6586]{P.~Hoeflich}
 \email{phoeflich77@gmail.com}
 \affiliation{Department of Physics, Florida State University, 77 Chieftan Way, Tallahassee, FL 32306, USA}

 \author[0000-0001-5393-1608]{E.~Baron}
  \email{ebaron@psi.edu}
 \affiliation{Planetary Science Institute, 1700 East Fort Lowell Road, Suite 106,
  Tucson, AZ 85719-2395, USA}
 \affiliation{Hamburger Sternwarte, Gojenbergsweg 112, D-21029 Hamburg, Germany}

\author[0000-0002-9301-5302]{M.~Shahbandeh}
 \email{mshahbandeh@stsci.edu}
\affiliation{Space Telescope Science Institute, 3700 San Martin Drive, Baltimore, MD 21218-2410, USA}

\author[0000-0002-7566-6080]{J.~M.~DerKacy}
 \email{jmderkacy@gmail.com}
\affiliation{Space Telescope Science Institute, 3700 San Martin Drive, Baltimore, MD 21218-2410, USA}

\author[0009-0001-9148-8421]{E.~Fereidouni}
\email{ef22g@fsu.edu}
 \affiliation{Department of Physics, Florida State University, 77 Chieftan Way, Tallahassee, FL 32306, USA}

\author[0000-0002-7305-8321]{C.~M.~Pfeffer}
\email{cmpfeffer@vt.edu}
 \affiliation{Institute for Astronomy, University of Hawaiʻi at M\=anoa, 2680 Woodlawn Dr., Hawaiʻi, HI 96822, USA }

\author[0000-0001-6107-0887]{S. Shiber}
 \email{ss24dw@fsu.edu}
 \affiliation{Department of Physics, Florida State University, 77 Chieftan Way, Tallahassee, FL 32306, USA}
\email{sshiber1@lsu.edu}

\author[0000-0001-6272-5507]{P.~Brown}
 \email{pbrown@physics.tamu.edu}
\affiliation{George P. and Cynthia Woods Mitchell Institute for Fundamental Physics and Astronomy, Texas A\&M University, Department of Physics and Astronomy, College Station, TX 77843, USA}

\author[0000-0003-4625-6629]{C.~Burns}
 \email{cburns@carnegiescience.edu}
\affiliation{Observatories of the Carnegie Institution for Science, 813 Santa Barbara Street, Pasadena, CA 91101, USA}

\author[0000-0001-7101-9831]{A.~Cikota}
 \email{aleksandar.cikota@noirlab.edu}
\affiliation{Gemini Observatory/NSF's NOIRLab, Casilla 603, La Serena, Chile}

\author[0000-0001-6069-1139]{T.~de~Jaeger}
 \email{thomas.dejaeger@lpnhe.in2p3.fr}
\affiliation{LPNHE, (CNRS/IN2P3, Sorbonne Universit\'{e}, Universit\'{e} Paris Cit\'{e}), Laboratoire de Physique Nucl\'{e}aire et de Hautes \'{E}nergies, 75005, Paris, France}

\author[0000-0003-3429-7845]{A.~Do}
 \email{ajmd6@cam.ac.uk}
\affiliation{Institute of Astronomy and Kavli Institute for Cosmology, University of Cambridge, Madingley Road, Cambridge CB3 0HA, UK}

\author[0000-0002-6230-0151]{D.~O.~Jones}
 \email{dojones@hawaii.edu}
 \affiliation{Institute for Astronomy, University of Hawaiʻi at M\=anoa, 2680 Woodlawn Dr., Hawaiʻi, HI 96822, USA }

\author[0000-0002-1296-6887]{L.~Galbany}
 \email{lluisgalbany@gmail.com}
\affiliation{Institute of Space Sciences (ICE, CSIC), Campus UAB, Carrer de Can Magrans, s/n, E-08193 Barcelona, Spain}
\affiliation{Institut d’Estudis Espacials de Catalunya (IEEC), E-08034 Barcelona, Spain}

\author[0000-0003-3953-9532]{W.~B.~Hoogendam}
\email{willemh@hawaii.edu} 
 \affiliation{Institute for Astronomy, University of Hawaiʻi at M\=anoa, 2680 Woodlawn Dr., Hawaiʻi, HI 96822, USA }

\author[0000-0003-1039-2928]{E.~Hsiao}
 \email{yichi.hsiao@gmail.com}
\affiliation{Department of Physics, Florida State University, 77 Chieftan Way, Tallahassee, FL 32306, USA}

\author[0000-0002-6650-694X]{K.~Krisciunas}
 \email{krisciunas@physics.tamu.edu}
\affiliation{George P. and Cynthia Woods Mitchell Institute for Fundamental Physics and Astronomy, Texas A\&M University, Department of Physics and Astronomy, College Station, TX 77843, USA}

\author[0000-0001-8367-7591]{S.~Kumar}
 \email{sahanak@gmail.com}
\affiliation{Department of Astronomy, University of Virginia, 530 McCormick Rd., Charlottesville, VA 22904, US}

\author[0000-0002-3900-1452]{J.~Lu}
 \email{lujing8@msu.edu}
\affiliation{Department of Physics and Astronomy, Michigan State University, East Lansing, MI 48824, USA}

\author[0000-0001-6876-8284]{P.~Mazzali}
 \email{P.Mazzali@ljmu.ac.uk}
\affiliation{Astrophysics Research Institute, Liverpool John Moores University, UK}
\affiliation{Max-Planck Institute for Astrophysics, Garching, Germany}

\author[0000-0003-2535-3091]{N.~Morrell}
 \email{nmorrell@carnegiescience.edu}
\affiliation{Las Campanas Observatory, Carnegie Observatories, Casilla 601, La Serena, Chile}

\author[0000-0003-2734-0796]{M.~Phillips}
 \email{mmp@lco.cl}
\affiliation{Las Campanas Observatory, Carnegie Observatories, Casilla 601, La Serena, Chile}

\author[0000-0003-4631-1149]{B.~Shappee}
 \email{shappee@hawaii.edu}
 \affiliation{Institute for Astronomy, University of Hawaiʻi at M\=anoa, 2680 Woodlawn Dr., Hawaiʻi, HI 96822, USA }

\author[0000-0002-5571-1833]{M. D.~Stritzinger}
 \email{max@phys.au.dk}
\affiliation{Department of Physics and Astronomy, Aarhus University, Ny Munkegade 120, DK-8000 Aarhus C, Denmark}

\author[0000-0002-8102-181X]{N.~Suntzeff}
 \email{nsuntzeff@tamu.edu}
\affiliation{George P. and Cynthia Woods Mitchell Institute for Fundamental Physics and Astronomy, Texas A\&M University, Department of Physics and Astronomy, College Station, TX 77843, USA}

\author[0000-0002-2471-8442]{M.~Tucker}
 \email{tuckerma95@gmail.com}
\altaffiliation{CCAPP Fellow}
\affiliation{Center for Cosmology and AstroParticle Physics, The Ohio State University, 191 W. Woodruff Ave., Columbus, OH 43210, USA}

\author[0000-0001-7092-9374]{L.~Wang}
 \email{qnwang@mit.edu}
\affiliation{Department of Physics and Astronomy, Texas A\&M University, College Station, TX 77843, USA}

\author[0000-0002-6535-8500]{Y.~Yang}
\email{yiyangtamu@gmail.com}
\affiliation{Department of Physics, Tsinghua University, Qinghua Yuan, Beijing 100084, China}